\documentclass[prd,aps,floats,floatfix,eqsecnum,nofootinbib]{revtex4}
\usepackage{amsmath,amssymb,verbatim,graphicx,rotating}
\usepackage{bm}

\newcommand{\be}{\begin{equation}}
\newcommand{\ee}{\end{equation}}
\newcommand{\bea}{\begin{eqnarray}}
\newcommand{\eea}{\end{eqnarray}}
\newcommand{\bds}[1]{\boldsymbol{#1}}
\newcommand{\vx}{\bds x}
\newcommand{\vk}{\bds k}

\begin{document}
\title{The CMB Quadrupole depression produced by early
fast-roll inflation: \\ MCMC analysis of WMAP and SDSS data.}
\author{\bf C. Destri$^{(a)}$} \email{Claudio.Destri@mib.infn.it}
\author{\bf H. J. de Vega$^{(b,c)}$} \email{devega@lpthe.jussieu.fr}
\author{\bf N. G. Sanchez$^{(c)}$} \email{Norma.Sanchez@obspm.fr}
\affiliation{$^{(a)}$ Dipartimento di Fisica G. Occhialini, Universit\`a
Milano-Bicocca Piazza della Scienza 3, 20126 Milano and
INFN, sezione di Milano, via Celoria 16, Milano Italia\\
$^{(b)}$ LPTHE, Laboratoire Associ\'e au CNRS UMR 7589,\\
Universit\'e Pierre et Marie Curie (Paris VI) et Denis Diderot (Paris VII),\\
Tour 24, 5 \`eme. \'etage, 4, Place Jussieu, 75252 Paris, Cedex 05, France.\\
$^{(c)}$ Observatoire de Paris, LERMA, Laboratoire Associ\'e au CNRS UMR 8112,
 \\61, Avenue de l'Observatoire, 75014 Paris, France.}
\begin{abstract}
  Generically, the classical evolution of the inflaton has a brief {\bf fast
    roll stage} that precedes the slow roll regime. The fast roll stage leads to
  a purely attractive potential in the wave equations of curvature and tensor
  perturbations (while the potential is purely repulsive in the slow roll
  stage). This attractive potential leads to a {\bf depression} of the CMB
  quadrupole moment for the curvature and B-mode angular power spectra.  A
  single {\bf new} parameter emerges in this way in the early universe model:
  the comoving wave number $ k_1 $ characteristic scale of this attractive
  potential.  This mode $ k_1 $ happens to exit the horizon precisely {\bf at
    the transition} from the fast-roll to the slow-roll stage.  The fast-roll
  stage dynamically modifies the initial power spectrum by a transfer function $
  D(k) $. We compute $ D(k) $ by solving the inflaton evolution equations. $
  D(k) $ effectively suppresses the primordial power for $ k < k_1 $ and
  posseses the scaling property $ D(k) = \Psi(k/k_1) $ where $ \Psi(x) $ is an
  universal function. We perform a MCMC analysis of the WMAP and SDSS data
  including the fast-roll stage and find the value $ k_1 = 0.266 \; {\rm
    Gpc}^{-1} $.  The quadrupole mode $ k_Q = 0.242 \; {\rm Gpc}^{-1} $ exits
  the horizon earlier than $ k_1 $, about one-tenth of an efold before the end
  of fast-roll. We compare the fast-roll fit with a fit without fast roll but
  including a sharp lower cutoff on the primordial power.  Fast-roll provides a
  slightly better fit than a sharp cutoff for the TT, TE and EE modes.
  Moreover, our fits provide {\bf     non-zero} lower bounds for $r$, 
while the values of the other cosmological
  parameters are essentially those of the pure $\Lambda$CDM model.  We display
  the real space two point $ C^{TT}(\theta) $ correlator. The fact that $ k_Q $ exits the
  horizon before the slow-roll stage implies an {\bf upper} bound in the total
  number of efolds $ N_{tot} $ during inflation.  Combining this with estimates
  during the radiation dominated era we obtain $ N_{tot} \sim 66 $, with the 
  bounds $ 62 < N_{tot} < 82 $.  We repeated the same analysis with the WMAP-5, 
ACBAR-2007 and SDSS data confirming the overall picture.
\end{abstract}

\date{\today}
\maketitle  
\tableofcontents

\section{Introduction and Results}

The Standard (Concordance) Model of the Universe
explains today a wide set of cosmological and astronomical measurements 
performed over a large variety of wave-lenghts and observation tools: 
large and small angular scale CMB observations, light elements abundances,
large scale structure observations (LSS) and properties of galaxy 
clusters, Hubble Space Telescope measurements on the Hubble constant, 
supernova luminosity/distance relations 
(acceleration of the today universe expansion), and other measurements. 
The concordance of these data imply that our universe is spatially flat, 
with gravity and cosmological perturbations described by 
Einstein General Relativity theory. 
WMAP data give a strong support to the Standard Model of the Universe.

Inflation was introduced to solve several outstanding problems of the
standard Big Bang model \cite{guthsato} and has now become an important 
part of the Standard Model of the Universe.
At the same time, it provides a natural mechanism
for the generation of scalar density fluctuations that seed large scale
structure, thus explaining the origin of the temperature anisotropies in
the cosmic microwave background (CMB), as well as that of tensor
perturbations (primordial gravitational waves) \cite{hu,mass}.

The horizon and flatness problems are solved provided
the universe expands for more than $ ~ 62 $ efolds during inflation.
This is achieved within slow-roll inflation where
the inflaton potential is fairly flat.

\medskip

Although there are no statistically significant departures from the
slow roll inflationary scenario at small angular scales ($ l\gtrsim
100 $), the WMAP data again confirms the surprinsingly
low quadrupoles $ C_2^{TT} $ and   $ C_2^{TE} $
\cite{WMAP-3}-\cite{WMAP-5} and suggests that
it cannot be completely explained by galactic foreground
contamination. The low value of the quadrupole has been an
intriguing feature on large angular scales since first observed by
COBE/DMR \cite{cobe}, and confirmed by the WMAP data \cite{WMAP-3}-\cite{WMAP-5}.

In order to asses the relevance of the observed quadrupole suppression
in the  $\Lambda$CDM model, we determine 
in the best fit $\Lambda$CDM model the probability to observe the 
quadrupole 20\% below the theoretical mean value.
This probability turns out to be only $ \sim 0.06 $. This small probability supports 
the necessity for a cosmological explanation of the quadrupole depression
beyond the $\Lambda$CDM model.

\medskip

Generically, the classical evolution of the inflaton has a brief {\bf
    fast roll stage} that precedes the slow roll regime. The fast roll
  stage leads to a purely attractive potential in the wave equations of
  curvature and tensor perturbations. 
Such potential is a \emph{generic} feature of this brief
\emph{fast roll} stage that merges smoothly with   slow roll
inflation. This stage is a consequence of generic initial conditions
\emph{for the classical inflaton dynamics} in which the kinetic and
potential energy of the inflaton are of the same order, namely, the
energy scale of slow roll inflation. During the early fast roll
stage the inflaton evolves rapidly during a brief period,  but slows
down by the cosmological expansion settling in  the slow roll stage
in which the kinetic energy of the inflaton is much smaller than its
potential energy.

As shown in ref. \cite{quadru1,quadru2} the attractive potential in the wave equations of
curvature and tensor perturbations during the fast-roll stage 
leads to a {\bf suppression} of the
quadrupole moment for CMB and B-mode angular power spectra.
Both scalar and tensor low multipoles are suppressed. However,
the potential for tensor perturbations is about an order of magnitude
smaller than the one for scalar fluctuations and hence the suppression
of  low $ \ell $ tensor perturbations is much less significative \cite{quadru1,quadru2}.

The observation of a low quadrupole \cite{cobe,WMAP-3,WMAP-5} 
and the surprising alignment of quadrupole
and octupole \cite{tegmark,virgo} sparked many different proposals
for their explanation \cite{expla}.

The fast-roll explanation of the quadrupole does not require to introduce new
physics neither modifications of the slow-roll inflationary models.
The only new feature is that the quadrupole mode should exit the horizon during
the generic fast-roll stage that precedes slow-roll inflation.

A single {\bf new} parameter emerges dynamically due to the fast-roll stage: the
comoving wave number $ k_1 $, characteristic scale of the attractive potential
felt by the fluctuations during fast-roll.  The fast-roll stage modifies the
initial power spectrum by a transfer function $ D(k) $ that we compute solving
the classical inflaton evolution equations [see fig. 3]. $ D(k) $ effectively
suppresses the primordial power for $ k < k_1 $ and posseses the scaling
property $ D(k) = \Psi(k/k_1) $ where $ \Psi(x) $ is an universal function. $
D(k) $ has a main peak around $ k_M \simeq 1.9 \; k_1 $ and oscillates around
zero with decreasing amplitude as a function of $ k $ for $ k > k_M $.  $ D(k) $
vanishes asymptotically for large $ k $, as expected.

\medskip

We report here the results of a MCMC analysis of the WMAP-3, small--scale CMB
and SDSS data including the fast-roll stage and find the value $ k_1 = 0.266 \;
{\rm Gpc}^{-1} $. This mode $ k_1 $ happens to exit the horizon precisely at the
transition from the fast-roll to the slow-roll stage.  The quadrupole mode $ k_Q
= 0.242 \; {\rm Gpc}^{-1} $ exits the horizon {\bf during} the fast-roll stage
approximately {\bf 1$/$10 of an efold earlier} 
than $ k_1 $.  We compare the
fast-roll fit with a fit without fast roll but including a sharp lower cutoff on
the primordial power.  Fast-roll provides a  slightly
better fit than a sharp cutoff 
for the $ C_{\ell}^{\rm TT} , \;  C_{\ell}^{\rm TE} $
and $ C_{\ell}^{\rm EE} $ coefficients.
Besides reproducing the quadrupole depression, the fast roll fit accounts for
the oscillations of the lower multipole data.

\medskip

We analyze with MCMC and compare three classes of cosmological models:

\begin{itemize}
\item The usual slow-roll $\Lambda$CDM, the $\Lambda$CDM$+r$ and the
  $\Lambda$CDM$+r$ on $ C_\mathrm{BNI} $ models.  BNI stands for {\em Binomial
    New Inflation}. In this last model we {\bf enforce} the theoretical
  functional relation (denoted $ C_\mathrm{BNI} $) between $ n_s $ and $ r $
  valid in BNI.(We call $\Lambda$CDM$+r$ on $ C_\mathrm{BNI} $ the usual
  $\Lambda$CDM$+r$ model constrained on the curve $ C_\mathrm{BNI} $).

\item The slow-roll $\Lambda$CDM on $ C_\mathrm{BNI} $ model with a sharp cut
  for $ k < k_1 $.
\item The $\Lambda$CDM on $ C_\mathrm{BNI} $ model including both fast and
  slow-roll stages.
\end{itemize}

The MCMC analysis of the WMAP and SDSS data favours a double-well, spontaneously
broken symmetric potential for the inflaton in new inflation \cite{mcmc}
$$
V(\varphi) = \frac{\lambda}4 \left( \varphi^2- \frac{m^2}{\lambda} \right)^2 \; .
$$
The quartic coupling in the effective theory of inflation \cite{1sN} is given by
$$ 
\lambda = \frac{y}{8 \; N} \left( \frac{M}{M_{Pl}}\right)^4 \sim 10^{-12} \; .
$$ 
Here $ N \sim 60 $ is the number of efolds since the 
cosmologically relevant modes exit the horizon till the end of inflation
and $ M_{Pl} = 2.4 \times 10^{18} $ GeV is the Planck mass. MCMC yields
for the dimensionless quartic coupling $ y \simeq 1.32 $ and
$$
M = 0.57 \times 10^{16} \;  {\rm GeV} \quad ,  \quad
m = 1.34 \times 10^{13} \;  {\rm GeV}
$$
for the inflation energy scale $ M $ and the inflaton mass scale $ m $,
respectively.

\medskip

We modified the {\em CosmoMC} code introducing the fast roll transfer function $
D_\mathcal{R}(k) $ in the primordial power spectrum according to
eq.~(\ref{curvapot}). 

We repeated the same analysis with the WMAP-5, ACBAR-2007 and SDSS data,
this time setting $N=60$, with no statistically significant change.

\medskip

Our fits imposing $ C_\mathrm {BNI} $ predict {\bf non-zero} lower bounds on $ r
$: at 95\% CL, we find that $r>0.023$ when no cutoff is 
introduced, while $r>0.018$
when either the sharp cutoff or the fastroll $D(k)$ are introduced.  
The best
fit values of the other cosmological parameters remain practically unchanged as
compared to $\Lambda$CDM. Similarly their marginalized probability 
distributions
are almost unchanged, with the natural exception of $n_s$, which in BNI has a
theoretical upper limit [see eq.~(\ref{cotns})].

\medskip

We observe that the oscillatory form of the fastroll transfer function $
D_\mathcal{R}(k) $, by {\bf depressing as well as enhancing} the primordial
power spectrum at long wavelengths, leads also to new superimposed {\bf
  oscillatory corrections} on the low multipoles. As far as fitting to current
data is concerned, such corrections are more effective than the pure reduction
caused by a sharp cutoff. The fast-roll oscillations
yield better gains in likelihood than the sharpcut case.  

We display the best fit for the $ C_{\ell}^{\rm TT} , \; 
C_{\ell}^{\rm TE} $ and $ C_{\ell}^{\rm EE} $ multipoles compared to the
experimental data at low $ \ell $. 
One can observe that for $ \ell = 2 $ and $
\ell = 3 $ fast-roll and sharpcut models yield rather similar results (and
better than the $\Lambda$CDM$+r$ model) while for $ \ell = 4 $ fast-roll
produces for $ C_{\ell}^{\rm TE} $ a value closer to WMAP-3 data than sharpcut. 
For $ C_{\ell}^{\rm EE} $ both fastroll and sharpcut models produce a depression
of the low multipoles including the EE quadrupole.

\medskip

We summarize in the Appendix the numerical code used by us in the simulations.

\medskip

We display the real space two point TT-correlator $ C^{TT}(\theta) $ for purely slow-roll
$\Lambda$CDM, sharpcut and fast-roll $\Lambda$CDM models. 
The purely slow roll $\Lambda$CDM correlator differs from the two others only for 
large angles $ \theta \gtrsim 1 $.
Since all $l$-modes besides the lowest ones are practically identical in the three cases, 
this shows how important are the low multipoles in the large angle correlations.

\medskip

We get the following picture of the inflationary universe 
explaining the quadrupole suppression from the effective 
(Ginsburg-Landau) theory of inflation combined with MCMC simulations 
of CMB$+$LSS data. A fast-roll stage lasting 
about one efold is followed by a slow-roll stage lasting $ \sim 65 $ 
efolds. We have the radiation dominated era after these $ \sim 65+1 = 66 $
efolds of inflation. The quadrupole modes exit the horizon during
the fast-roll stage about $ 0.4 $ of an efold after the beginning of inflation
and is therefore {\bf suppressed} compared with the modes exiting later
the horizon during the slow-roll stage.

\medskip

The fast-roll stage explains the quadrupole suppression and {\bf fixes the total
  number of efolds} of inflation. The fact that the quadrupole mode $ k_Q $
exits the horizon before the slow-roll stage implies an {\bf upper} bound in the
total number of efolds $ N_{tot} $ during inflation. Combining this with
estimates during the radiation dominated era we obtain $ N > 56 $, 
$ N_{tot}\sim 66 $, the upper bound $ N_{tot} < 82 $ and the lower 
bound $ N_{tot}>62 $.

\medskip

Our MCMC simulations give good fits for $ N=50 $ and $ N=60 $. The bound $ N > 56
$ therefore favours $ N \sim 60 $ which implies $ N_{tot} \sim 66 $ and $ H \sim
3 \times 10^{10} $ GeV by the end of inflation.

Changing $ N $ from 50 to 60 does not affect significatively the MCMC fits we
present in this paper. This is partially due to the fact that a change on $ y $
can partially compensate a change on $ N $. More importantly, a 20\% change in
$N$ may affect the fit of $k_1$ by a similar amount, 
leaving unchanged its
scale, which is of the order of the inverse Hubble scale.

Another {\bf hint} to increase $ N $ above 50 comes from WMAP-5 that gives a
larger $ n_s $ value and using the theoretical upper limit for $ n_s $
\cite{prd,mcmc}: $ n_s < 1 - \frac{1.9236\ldots}{N} $, which gives $ n_s <
0.9679 \ldots $ for $ N = 60 $. This value is compatible with the $ n_s $ value
from WMAP$5+$BAO$+$SN and no running \cite{WMAP-5}.

\section{The effective Theory of Inflation. Fast and slow roll regimes.} \label{potuniv}

The inflaton potential $ V(\varphi) $ must be a slowly varying
function of $ \varphi $ in order to permit a slow-roll solution for
the inflaton field which guarantees a large enough total number of efolds $ \gtrsim 62 $.
Such value is necessary to solve the horizon, flatness and entropy problems.

As discussed in ref. \cite{1sN}, the inflaton potential should have the universal form
\be \label{V} 
V(\varphi) = N \; M^4 \; w(\chi)  \; ,
\ee  
\noindent where $ \chi $ is a dimensionless, slowly varying field 
\be\label{chifla} 
\chi = \frac{\varphi}{\sqrt{N} \;  M_{Pl}}  \; ,
\ee 
and $ M $ is the energy scale of inflation, $ N \sim 60 $ is the number of efolds since the 
cosmologically relevant modes exit the horizon till the end of inflation.

The energy scale $ M $ of inflation is determined by the amplitude of the observed 
CMB anisotropy, which implies  $ M \sim 0.7 \times 10^{16} $ GeV. That is, $  M \ll  M_{Pl} $,
which ensures the consistency of the effective theory of inflation.

\medskip

The dynamics of the rescaled field $ \chi $ exhibits the slow
time evolution in terms of the \emph{stretched} dimensionless time variable, 
\be \label{tau} 
\tau =  \frac{t \; M^2}{M_{Pl} \; \sqrt{N}}  \quad , \quad  
{\cal H} \equiv \frac{H \; M_{Pl}}{\sqrt{N} \; M^2} = {\cal O}(1) \; .
\ee 
The rescaled variables $ \chi $ and $ \tau $ change slowly with time. 
A large change in the field amplitude $ \varphi $ results in a small change 
in the $ \chi $ amplitude, a change in $ \varphi \sim  M_{Pl} $ results in a 
$ \chi $ change $ \sim 1/\sqrt{N} $. The form of the potential, eq.(\ref{V}),
the rescaled dimensionless inflaton field eq.(\ref{chifla}) and the time 
variable $ \tau $ make {\bf manifest} the slow-roll expansion as a consistent 
systematic expansion in powers of $ 1/N $ \cite{1sN}.  

\medskip

We can choose $ |w''(0)| = 1 $ without loosing generality. Then, 
the inflaton mass scale at zero field is given by a see-saw formula
\be \label{m}
m^2  = \left| V''(\varphi=0) \right| = 
\frac{M^4}{M_{Pl}^2}  \quad ,  \quad  M \sim 0.7 \times 10^{16} \, \textrm{GeV} \quad ,  \quad
m = \frac{M^2}{M_{Pl}} 
\sim 2.0 \times 10^{13} \, \textrm{GeV} \; .
\ee
The Hubble parameter when the cosmologically relevant modes exit the horizon
is given by
\be \label{Hi}
H  = \sqrt{N} \; m \, {\cal H} \sim 5 \; m 
\sim 1.0 \times 10^{14}\,\textrm{GeV}\; ,  
\ee
where $  {\cal H} \sim 1 $. As a result, $ m\ll M $ and 
$ H \ll M_{Pl} $. 

\medskip

The energy density and the pressure in terms of the dimensionless 
rescaled field $ \chi $ and the slow time variable $ \tau $ take the form,
\be\label{enepre2} 
\frac{\rho}{ N \; M^4} = \frac1{2\;N} \left(\frac{d\chi}{d \tau}\right)^2 + w(\chi) 
\quad ,\quad 
\frac{p}{ N \; M^4} = \frac1{2\;N} \left(\frac{d\chi}{d \tau}\right)^2 - w(\chi) \; . 
\ee
The equations of motion in the same dimensionless variables become
\bea \label{evol} 
&&  {\cal H}^2(\tau) = \frac13\left[\frac1{2\;N} 
\left(\frac{d\chi}{d \tau}\right)^2 + w(\chi) \right] \quad , \cr \cr
&& \frac1{N} \;  \frac{d^2
\chi}{d \tau^2} + 3 \;  {\cal H} \; \frac{d\chi}{d \tau} + w'(\chi) = 0 \quad .
\eea 
The slow-roll approximation follows by neglecting the
$ \frac1{N} $ terms in eqs.(\ref{evol}). Both
$ w(\chi) $ and $ {\cal H}(\tau) $ are of order $ N^0 $ for large $ N $. Both
equations make manifest the slow roll expansion as an expansion in $ 1/N $.

\medskip

The number of e-folds $ N[\chi] $ since the field $ \chi $ exits the horizon 
till the end of inflation (where $ \chi $ takes the value $ \chi_{end} $) 
can be computed in close form from eqs. (\ref{evol}) in the slow-roll 
approximation (that is, neglecting $ 1/N $ corrections):
\be \label{Ncho}
\frac{N[\chi]}{N} = -\int_{\chi}^{\chi_{end}}  \;
\frac{w(\chi)}{w'(\chi)} \; d\chi \;  \leqslant 1 \; ,
\ee
where we choose  $ N = N[\chi] $.
Therefore, eq.(\ref{Ncho}) determines $ \chi $ at horizon exit as a function of the
couplings in the inflaton potential $ w(\chi) $:
\be \label{Nchi}
-\int_{\chi}^{\chi_{end}}  \; \frac{w(\chi)}{w'(\chi)} \; d\chi = 1 \; .
\ee
Inflation ends after a finite number of efolds provided 
\be\label{condw}
w(\chi_{end}) = w'(\chi_{end}) = 0 \; .
\ee
So, this condition is enforced in the inflationary potentials.

\medskip

There are two {\it generic} inflationary regimes: slow-roll and fast-roll depending on
whether \cite{quadru2}
\bea
&& \frac1{2\;N} \left(\frac{d\chi}{d \tau}\right)^2 \ll w(\chi) \quad : \quad 
{\rm  slow-roll \; regime} \cr \cr
&& \frac1{2\;N} \left(\frac{d\chi}{d \tau}\right)^2 \sim w(\chi) \quad : \quad 
{\rm  fast-roll \; regime} \; .
\eea
Both regimes appear in {\bf all} inflationary models in the class eq.(\ref{V}).
Fast-roll clearly corresponds to generic initial conditions for the inflaton field.
The fast-roll stage turns to be very short and is generically followed by
the slow-roll stage \cite{quadru2}.

\medskip

For the quartic degree potentials $ V(\varphi) $, the main two families are:

\medskip

\noindent
(a) discrete symmetry ($ \varphi \to - \varphi $) breaking potentials 
(so-called new, or small-field, inflation)
\begin{equation}\label{nueva}
  V(\varphi) = \frac{\lambda}4 \left( \varphi^2- \frac{m^2}{\lambda} \right)^2 =
  -\frac{m^2}2 \; \varphi^2 +\frac{\lambda}4  \; \varphi^4 + 
  \frac{m^4}{4 \; \lambda} \; ;
\end{equation}
(b) unbroken symmetry potentials (chaotic, or large-field, inflation), 
\begin{equation}\label{caotica}
  V(\varphi)= +\frac{m^2}2  \; \varphi^2 +\frac{\lambda}4  \; \varphi^4 \; .
\end{equation}
The corresponding dimensionless potentials $ w(\chi) $ take the form
\begin{equation}\label{wnue}
  w(\chi) = \frac{y}{32} \left(\chi^2 - \frac8{y}\right)^{\! 2} = -\frac12 \; \chi^2 
  + \frac{y}{32} \; \chi^4 + \frac2{y} ~~~~~~ \text{for new inflation} 
\end{equation}
and
\begin{equation}\label{wchao}
  w(\chi) = \frac12 \; \chi^2 + \frac{y}{32} \; \chi^4  ~~~~~~ \text{for chaotic inflation}
\end{equation}
where the coupling $ y $ is of {\bf order one} and
$$
\lambda = \frac{y}{8 \; N} \left( \frac{M}{M_{Pl}}\right)^4 \sim 10^{-12} \quad .
$$
In new inflation the inflaton starts near the local maximum $ \chi = 0 $
and keeps rolling down the potential hill till it reaches the absolute minimum
$ \chi = \sqrt{\frac8{y}} $. In general, the initial kinetic energy may be of the
same order of magnitude as the initial potential energy of the inflaton which defines
fast-roll inflation. That is, in general the initial states are not slow-roll.

By numerically solving  eqs. (\ref{evol}) we find that the fast-roll initial stage of
the inflaton becomes very soon a slow-roll stage \cite{quadru2}.
This is a general property and implies that the slow-roll regime is
an atractor for this dynamical system \cite{bgzk}.
We see a de Sitter-like expansion during the slow-roll stage $ \tau \lesssim 3 $ during which
the Hubble parameter decreases slowly and monotonically.

An initial state for the \emph{inflaton} (inflaton classical
dynamics) with approximate \emph{equipartition} between kinetic and
potential energies is a more \emph{general} initialization of
cosmological dynamics in the effective field theory than slow roll
which requires that the inflaton kinetic energy is much smaller
than its potential energy. The most
\emph{generic} initialization of the inflaton dynamics in the
effective field theory leads to a \emph{fast roll} stage followed by
slow roll inflation \cite{quadru2}.

The total number of efolds of inflation is determined by the initial
conditions for the inflaton field: $ \chi(0) \sim {\dot\chi}(0)= {\cal O}(1) $.
Varying these initial conditions the total number of efolds of inflation
sweeps a wide range of efold values showing the flexibility of the inflationary model.

\medskip

We have carried out analogous numerical studies in scenarios of
chaotic inflation with  similar results: if the initial kinetic
energy of the inflaton is of the same order as the potential energy,
a \emph{fast roll} stage is \emph{always} present. The evolution
of the potentials $ {\mathcal V}_\mathcal{R}(\eta)$ and  $ {\mathcal V}_T(\eta)$
felt by the curvature and tensor perturbations are
similar to those for new inflation and they are always
\emph{attractive} during the fast roll stage (see below).

\begin{figure}[h]
  \includegraphics[scale=0.8]{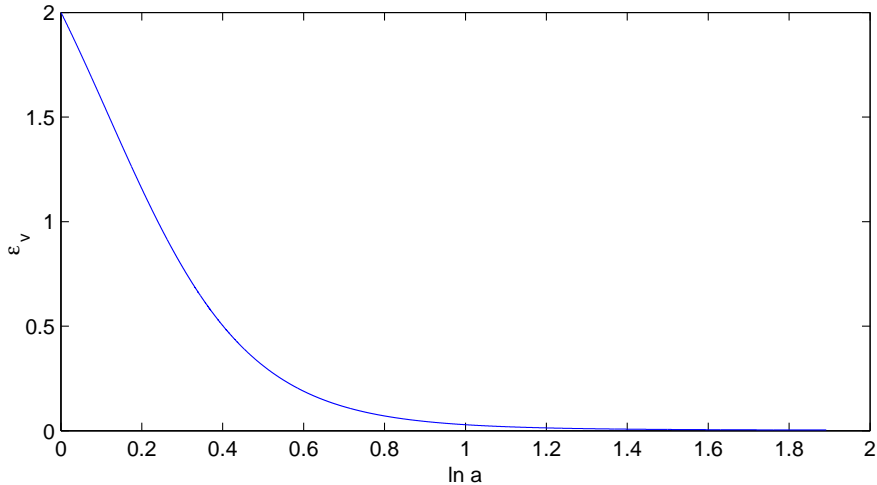}
  \caption{We plot here $ \epsilon_v $ vs. $ \ln a $ during the fast roll
    stage and the beginning of slow-roll for new inflation with $ y = 1.322
    $.  We define as the end of fast-roll the point where $ \epsilon_v =
    \frac1{N} = 0.02 $.  This gives here $ \ln a_F = 1.091 $. Namely,
    fast-roll ends one efold after the beginning of inflation.} 
  \label{ya}
\end{figure}

\section{The Effect of Fast-roll on the Inflationary Fluctuations.}

The inflationary scenario features scalar curvature fluctuations determined by a gauge
invariant combination of the inflaton field and metric fluctuations. They also
feature tensor fluctuations (gravitational waves).

It is convenient to  introduce the gauge invariant potential \cite{hu},
\begin{equation}\label{u}
  u(\vx,t)=-z \;  \mathcal{R}(\vx,t) \; , 
\end{equation}
where $ \mathcal{R}(\vx,t) $ stands for the gauge invariant curvature perturbation of the comoving
hypersurfaces and
\be \label{za} 
z \equiv a(t) \; \frac{\dot{\varphi} }{H} \;  .
\ee 
The gauge invariant curvature field $ u(\vx,t) $ expanded in terms of conformal time mode functions and
creation and annihilation operators take the form \cite{hu}
\be \label{curvau} 
u(\vx,\eta)=\int \frac{d^3k}{(2 \; \pi)^\frac32}\left[
\alpha_\mathcal{R}(\vk) \; S_\mathcal{R}(k;\eta) \; e^{i\vk\cdot\vx}  +
\alpha^\dagger_\mathcal{R}(\vk) \; S^*_\mathcal{R}(k;\eta)  \; e^{-i\vk\cdot\vx}\right] \; ,
\ee 
where the operators obey canonical commutation relations
\begin{equation*}
  \left[ \alpha_\mathcal{R}(\vk), \; \alpha^\dagger_\mathcal{R}(\vk') \right]
  = \delta^{(3)}(\vk-\vk') \; .
\end{equation*}
The vacuum state is annihilated by the operators $ \alpha_\mathcal{R}(k) $ and
the mode functions obey the equations of motion \cite{hu},
\begin{equation}\label{fluces}
  \Bigg[\frac{d^2}{d\eta^2}+k^2- \frac1{z} \frac{d^2z}{d\eta^2}
  \Bigg]S_\mathcal{R}(k;\eta) =0 \,. 
\end{equation}
Here $ \eta $ stands for the conformal time
\begin{equation}\label{confo}
  \eta = \int \frac{dt}{a(t)} \; .
\end{equation}
Eq. (\ref{fluces}) is a Schr\"odinger-type differential equation in the variable
$ \eta $.  The potential felt by the fluctuations
\begin{equation}\label{WC}
  W_\mathcal{R}(\eta) \equiv \frac1{z} \; \frac{d^2 z}{d \eta^2}
\end{equation}
can be expressed in terms of the inflaton potential and its derivatives. From
eqs.(\ref{za}) and (\ref{WC}) and using the inflation equations of motion
(\ref{evol}), the potential $ W_\mathcal{R}(\eta) $ can be written as
\cite{quadru2}
\begin{equation}\label{W} 
  W_\mathcal{R}(\eta)  = a^2(\eta) \; H^2(\eta) \left[ 2 - 7 \, \epsilon_v
    + 2 \, \epsilon_v^2 - \frac{\sqrt{8 \; \epsilon_v} \; V'}{M_{Pl} \;
      H^2} - \eta_v (3 -  \epsilon_v) \right] \; , 
\end{equation}
where we take for the sign of the square root $ \sqrt{\epsilon_v} $ the
sign of $ \dot{\varphi} $ and
\begin{equation}\label{srp}
  \epsilon_v \equiv  \frac1{2 \; M_{Pl}^2} \; \frac{\dot{\varphi}^2}{H^2} \quad ,\quad
  \eta_v \equiv  M_{Pl}^2 \; \frac{V''(\varphi)}{V(\varphi)} \; .
\end{equation}
$ \epsilon_v $ and $ \eta_v $ are the known slow-roll parameters \cite{hu}.
Notice that eqs. (\ref{W})-(\ref{srp}) are {\bf exact} (no slow-roll approximation).

\medskip

In terms of the dimensionless variables eqs.(\ref{V})-(\ref{tau}) we obtain for
the potential $ W_\mathcal{R}(\eta) $,
\begin{equation}\label{Wb}
  W_\mathcal{R}(\eta) =  a^2(\eta) \; {\cal H}^2 \; m^2 \; N \left[ 2 - 7 \;
    \epsilon_v + 2 \; \epsilon_v^2 - \sqrt{\frac{8 \; \epsilon_v }{N}}
    \frac{w'}{{\cal H}^2}  - \eta_v (3 -  \epsilon_v)\right] \;, 
\end{equation}
while the parameters $ \epsilon_v $ and $ \eta_v $ take the form 
\be\label{slrsd}
\epsilon_v = \frac1{2 \; N} \; \frac1{{\cal H}^2} \; \left( \frac{d \chi}{d \tau}\right)^2 
\quad , \quad   \eta_v = \frac1{N}  \; \frac{w''(\chi)}{w(\chi)} \; .
\ee
In the slow-roll regime they can be approximated as
\be
\epsilon_v = \frac1{2 \; N} \; \left[\frac{w'(\chi)}{w(\chi)} \right]^2 +{\cal O}\left(\frac1{N^2}\right)
= {\cal O}\left(\frac1{N}\right)  
\quad , \quad   \eta_v = \frac1{N}  \; \frac{w''(\chi)}{w(\chi)} = {\cal O}\left(\frac1{N}\right) \; .
\ee
We explicitly see that the parameters  $ \epsilon_v $ and $ \eta_v $ are suppressed by powers of 
$ 1/N $ in the slow-roll regime.
This result is valid for {\bf all} models in the class defined by eq.(\ref{V}) 
regardless of the precise form of $ w(\chi) $. 

\bigskip

Tensor perturbations (gravitational waves) are gauge invariant.
The corresponding quantum fields (gravitons) are written as 
\begin{equation}\label{tensh}
  h^i_j(\vx,\eta) = \frac2{a(\eta) \; M_{Pl}}
  \sum_{\lambda=\times,+} \int \frac{d^3k}{(2 \; \pi)^\frac32}
  \; \epsilon^i_j(\lambda,\vk) \left[e^{i \vk \cdot \vx} \; 
    a_{\lambda,\vk} \; \,S_T(k,\eta)+e^{-i \vk\cdot \vx} \; 
    a^\dagger_{\lambda,\vk} \; \,S^*_T(k,\eta) \right] \; ,
\end{equation}
\noindent where $ \lambda $ labels the two standard
transverse and traceless polarizations $ \times $ and $ + $. The
operators $ \alpha_{\lambda,\vk}, \; \alpha^\dagger_{\lambda,\vk} $ 
obey canonical commutation relations, and $ \epsilon_{ij}(\lambda,\vk) $ are
the two independent symmetric and traceless-transverse tensors constructed from
the two independent polarization vectors transverse to $ \vk$, chosen to be real 
and normalized such that $ \epsilon^i_j(\lambda,\vk)\, \; 
\epsilon^j_k(\lambda',\vk)=\delta^i_k \; \delta_{\lambda,\lambda'} $.

\medskip

The mode functions $ S_T(k;\eta) $ obey the differential equation \cite{hu,quadru1,quadru2}
\be\label{Sten} 
S^{''}_{T}(k;\eta)+\left[k^2-
\frac{a''(\eta)}{a(\eta)}\right]S_{T}(k;\eta) = 0 \; . 
\ee

\medskip

That is, for both scalar curvature and tensor equations we have the equation
\be
\left[\frac{d^2}{d\eta^2}+k^2-W(\eta)\right]S(k;\eta) = 0 \; . \label{Wgen}
\ee
where for scalar curvature perturbations $ W_\mathcal{R}(\eta) $ is given by eq.(\ref{WC}) 
and for tensor perturbations $ W_T(\eta) $ is
$$ 
W_T(\eta) = \frac{a''(\eta)}{a(\eta)} \; .
$$
It is convenient to  explicitly separate the behavior of $ W(\eta) $
during the slow roll stage by writing
\be\label{defV}
W(\eta)=  \mathcal{V}(\eta) +\frac{\nu^2-\frac14}{\eta^2} \; ,
\ee
where the potential $ \mathcal{V}(\eta) $ is the fast-roll part and,
\be
\nu =  \Bigg\{ \begin{array}{l}
\nu_{\mathcal{R}}=\frac32+ 3 \, \epsilon_v -\eta_v +  {\cal O}\left(\frac1{N^2}\right)
\quad \mathrm{for~curvature~perturbations}        \\ \\
\nu_T = \frac32+ \epsilon_v + {\cal O}\left(\frac1{N^2}\right) \quad 
 \mathrm{for~tensor~perturbations} \; .  \\
\end{array} \label{defnu}
\ee
$ \epsilon_v $ and $ \eta_v $ are given by eqs.(\ref{srp}),(\ref{slrsd}).

\medskip

The potential $ \mathcal{V}(\eta) $ is localized in the fast roll stage \emph{prior} 
to slow roll (during which cosmologically relevant modes cross out of the Hubble radius),
$ \mathcal{V}(\eta) $ vanishes during slow-roll. In terms of the potential
$\mathcal{V}(\eta)$ the equations for the quantum fluctuations read,
\be 
\left[\frac{d^2}{d\eta^2}+k^2-\frac{\nu^2-\frac14}{\eta^2}-
\mathcal{V}(\eta)  \right]S(k;\eta) = 0 \; . \label{eqnpsr2} 
\ee

\subsection{The Fluctuations during the slow-roll stage}

The slow roll dynamics acts through the term 
$$ 
\frac{\nu^2-1/4}{\eta^2}
$$
which is a \emph{repulsive}  centrifugal barrier.

During the slow roll stage $ \mathcal{V}(\eta) $ is negligeable and the mode
equations simplify to 
\be \label{geneq} 
\left[\frac{d^2}{d\eta^2}+k^2 - \frac{\nu^2 -\frac14}{\eta^2} \right]S_{sr}(k,\eta) = 0 \; .
\ee 
To leading order in slow roll, $ \nu $ is constant
and for general initial conditions the solution of eq.(\ref{geneq}) is, 
\be \label{genS}
S_{sr} (k;\eta) = A (k) \; g_{\nu}(k;\eta) + B (k) \; [g_{\nu }(k;\eta)]^* \; ,
\ee 
where 
\be
g_{\nu }(k;\eta) =  \frac12 \; i^{\nu +\frac12} \;
\sqrt{-\pi \eta}\,H^{(1)}_{\nu }(-k\eta) \; , \label{gnu}
\ee
\noindent $ H^{(1)}_\nu(z) $ are Hankel functions. These
solutions are normalized so that their Wronskian is given by
\be\label{wronskian}
W[g_\nu(k;\eta),g^*_\nu(k;\eta)]=
g'_\nu(k;\eta) \; g^*_\nu(k;\eta)-g_\nu(k;\eta) \; [g'_\nu(k;\eta)]^* = -i \; .
\ee
The mode functions and coefficients $ A(k), \; B(k) $ will feature a subscript index
$ {\mathcal{R}}, \; T $, for curvature or tensor perturbations, respectively.

For wavevectors deep inside the Hubble radius $ | k \, \eta | \gg 1
$, the mode functions have the asymptotic behavior
\be
g_{\nu}(k;\eta) \buildrel{\eta \to
-\infty}\over= \frac1{\sqrt{2 \, k}} \; e^{-ik\eta} \; , \label{fnuasy}
\ee
while for $ \eta \to 0^- $ the mode functions behave as:
\be\label{geta0}
g_\nu(k;\eta)\buildrel{\eta \to 0^-}\over=
\frac{\Gamma(\nu)}{\sqrt{2 \, \pi \; k}} \; \left(\frac2{i \; k \;
\eta} \right)^{\nu - \frac12} \; .
\ee
In particular, in the scale invariant case $ \nu=\frac32 $ which is
the leading order in the slow roll expansion, the mode functions eqs.(\ref{gnu})
simplify to
\be
g_{\frac32}(k;\eta) =
\frac{e^{-ik\eta}}{\sqrt{2k}}\left[1-\frac{i}{k\eta}\right] \; . \label{g32}
\ee

\subsection{The Fluctuations during the earlier fast-roll stage}

The mode equation (\ref{eqnpsr2}) can be written as an integral equation.
We choose as initial condition the usual Bunch-Davies asymptotic condition
\be \label{BD}
S(k;\eta \rightarrow -\infty) = g_{\nu}(k;\eta \rightarrow -\infty) =
\frac{e^{-i \, k \, \eta}}{\sqrt{2k}} \; . 
\ee
We formally consider here inflation and the conformal time starting at $
\eta = -\infty $. However, it is natural to consider that the
inflationary evolution of the universe starts at some negative value
$ \eta_i < {\bar \eta} $, where $ \bar \eta $ is  the conformal
time when fast roll ends and slow roll begins.

The mode equation (\ref{eqnpsr2}) can be written as an integral equation
including the  Bunch-Davies initial condition eq.(\ref{BD}),
\be \label{solu}
S(k;\eta)= g_\nu(k;\eta) + i
\; g_\nu(k;\eta)\,\int^{\eta}_{-\infty}
 g^*_\nu(k;\eta') \; \mathcal{V}(\eta') \;  S(k;\eta') \; d\eta'-i  \;
g^*_\nu(k;\eta)\,\int^{\eta}_{-\infty}
 g_\nu(k;\eta') \; \mathcal{V}(\eta') \;  S(k;\eta') \; d\eta'\quad .
\ee
where for simplicity we set $ \eta_i = -\infty $.

Since $ \mathcal{V}(\eta) $ vanishes for  $ \eta > {\bar \eta} $,  the
mode functions $ S(k;\eta) $  for  $ \eta > {\bar \eta} $ can be written 
as linear combinations of the mode functions $ g_\nu(k;\eta) $ and $ g^*_\nu(k;\eta) $,
\be \label{solSR}
S(k;\eta) = A(k) \; g_\nu(k;\eta) + B(k) \; g^*_\nu(k;\eta)
\quad , \quad \eta > {\bar \eta} \quad ,
\ee
where the coefficients $ A(k) $ and $ B(k) $ can be read from eq.(\ref{solu}),
\bea
A(k) & = &  1+ i\int^{0}_{-\infty}
g^*_\nu(k;\eta) \; \mathcal{V}(\eta) \;  S(k;\eta) \;
d\eta\label{aofk} \\ \cr
 B(k) & = & -i \int^{0}_{-\infty}  g_\nu(k;\eta) \; \mathcal{V}(\eta)  \;
S(k;\eta) \;  d\eta\label{bofk}  \; .
\eea
The coefficients $ A(k) $ and $ B(k) $ are therefore {\bf calculated}
from the {\bf dynamics  before} slow roll, that is, during fast-roll.
[recall that $ \mathcal{V}(\eta)  = 0 $ for  $ \eta > {\bar \eta} $ during slow roll.]

\medskip

The constancy of the Wronskian
$ W[S(k;\eta),S^*(k;\eta)]=-i $ and eqs. (\ref{wronskian}), (\ref{solSR}) imply the
constraint,
$$
|A(k)|^2-|B(k)|^2=1 \quad .
$$
This relation permits to represent the  coefficients $ A(k); \; B(k) $ as \cite{quadru1}
\be
A(k) = \sqrt{1+N(k)} \; e^{i\theta_A(k)}~~;~~
B(k)=\sqrt{N(k)} \; e^{i\theta_{B}(k)} \label{bogonum} \; ,
\ee
\noindent where $ N(k), \; \theta_{A,{B}}(k) $ are real. 

\medskip

Starting with Bunch-Davies initial conditions for $ \eta \to
-\infty $, the action of the fast-roll potential $ \mathcal{V}(\eta) $ generates a mixture 
(Bogoliubov transformation) of the two linearly independent mode functions 
$ g_\nu(k;\eta) $ and $ g^*_\nu(k;\eta) $, which result in the mode
functions $ S(k;\eta) $ eq.(\ref{solSR}) for $ \eta > {\bar \eta} $ when the fast roll
potential $ \mathcal{V}(\eta) $ vanishes. 
This is clearly equivalent to starting the evolution
of the fluctuations at the \emph{beginning} of slow roll $ \eta =
{\bar \eta} $ with initial conditions defined by the Bogoliubov
coefficients $ A(k) $ and $ B(k) $ given by eq.(\ref{bofk}) as stressed in ref. \cite{quadru2}.

As shown in ref.\cite{quadru2} the power spectrum of curvature and tensor
perturbations for the general fluctuations eq.(\ref{solSR}) takes
the form, 
\bea\label{curvapot} 
&& P_\mathcal{R}(k) \buildrel{\eta \to  0^-}\over= \frac{k^3}{2 \; \pi^2} \; 
\Big|\frac{S_\mathcal{R}(k;\eta)}{z(\eta)} \Big|^2 =
P^{sr}_{\mathcal{R}}(k)\Big[1+D_\mathcal{R}(k) \Big] \; , \\ \cr 
&& P_T(k)  \buildrel{\eta \to 0^-}\over= \frac{k^3}{2 \; \pi^2} \;
\Big|\frac{S_T(k;\eta)}{C(\eta)} \Big|^2
= P^{sr}_T(k)\Big[1+ D_T(k) \Big] \; . \nonumber
\eea
Here $ D_\mathcal{R}(k) $ and $ D_T(k) $ are the transfer functions
for the initial conditions of curvature and tensor perturbations
introduced in ref.\cite{quadru1}: 
\be \label{DofkR} 
D(k) = 2 \; | B(k)|^2 -2 \; \mathrm{Re}\left[A(k)\; B^*(k)\,i^{2\nu-3}\right] = 2 \; N(k)-2
\; \sqrt{N(k)[1+N(k)]} \; \cos\left[
\theta_k - \pi \left(\nu - \frac32 \right) \right] \; . 
\ee 
where one uses either $ _\mathcal{R} $ or $ _T $ quantities and
$ \theta_k \equiv \theta_{B}(k)-\theta_A(k) $. 

\medskip

The standard slow roll power spectra are given by \cite{hu}: 

\bea  \label{potBD}
P^{sr}_\mathcal{R}(k) &=& \left( \frac{k}{2 \, k_0}\right)^{n_s - 1}
\; \frac{\Gamma^2(\nu)}{\pi^3} \; \frac{H^2}{2 \; \epsilon_v  \;
M_{Pl}^2 }\equiv \mathcal{A}^2_\mathcal{R} \;
\left(\frac{k}{k_0}\right)^{n_s - 1} \; ,\cr \cr P^{sr}_T(k) &=&
\mathcal{A}^2_T \; \left(\frac{k}{k_0}\right)^{n_T} \;   \quad ,
\quad n_T= -2 \; \epsilon_v \quad , \quad
\frac{\mathcal{A}^2_T}{\mathcal{A}^2_\mathcal{R}} = r = 16 \; \epsilon_v \; . 
\eea
These spectra are modified by the
fast-roll stage as displayed in eq.(\ref{curvapot}). The scale $k_0 $ is a
reference or pivot scale, for example WMAP takes $k_0 = 0.002\,{\rm Mpc}^{-1}$
and CosmoMC, $k_0 = 0.050\, {\rm Mpc}^{-1}$ (see section \ref{mcmcsec}).

\medskip

The integral equation (\ref{solu}) can be solved iteratively in a perturbative
expansion if the potential $ \mathcal{V}(\eta) $
is small when compared to 
$$ 
k^2 - \frac{\nu^2-1/4}{\eta^2} \; ,
$$
which is indeed true in this case.
Then, we can use for the coefficients $ A(k), \; B(k) $ the first
approximation obtained by replacing $ S (k;\eta') $ by
$ g_\nu(k;\eta') $ in the integrals eqs.(\ref{aofk})-(\ref{bofk}).
This is the Born approximation, in which
\be
A(k)   =    1+  i \int^{0}_{-\infty}
   \mathcal{V}(\eta)\,|g_\nu(k;\eta)|^2 \, d\eta   \quad  , \quad
B(k)  =  - i  \int^{0}_{-\infty}
   \mathcal{V}(\eta)\, g^2_\nu(k;\eta)  \, d\eta \; .\label{bofk0}
\ee
The transfer function of initial conditions given by eq.(\ref{DofkR}) can be computed
in the Born approximation, which is indeed appropriate in this situation. By using
eqs.(\ref{bofk0}) for the Bogoliubov coefficients $ A(k) $ and $ B(k) $ to dominant
order in $ 1/N $, that is $ \nu = 3/2 $ [eq.(\ref{defnu})],  $ D(k) $ is given by,
\be \label{Dborn2} 
D(k) = \frac1{k} \int^0_{-\infty} d\eta \;
\mathcal{V}(\eta) \left[\sin(2\, k \; \eta) \left(1  - \frac1{k^2 \,
\eta^2} \right)+ \frac2{k \, \eta} \, \cos(2\, k \, \eta) \right] \; .
\ee 
The potential $ \mathcal{V}(\eta) $ is obtained from eq.(\ref{defV}) as
$$
\mathcal{V}(\eta) = W(\eta) - \frac{\nu^2-1/4}{\eta^2} \; .
$$
To explicitly compute $ \mathcal{V}_\mathcal{R}(\eta) $ as a function of $
\eta $ for the curvature fluctuations we solve numerically the equations of
motion (\ref{evol}) for new inflation [eq.(\ref{wnue})] and insert the
solution for the inflaton $ \chi(\eta) $ in eqs.(\ref{Wb})-(\ref{slrsd}).
No large $ N $ approximation is used in this numerical calculation since we
cover in the evolution the fast-roll region where slow-roll obviously does
not apply.

\begin{figure}[h]
  \includegraphics[scale=0.8]{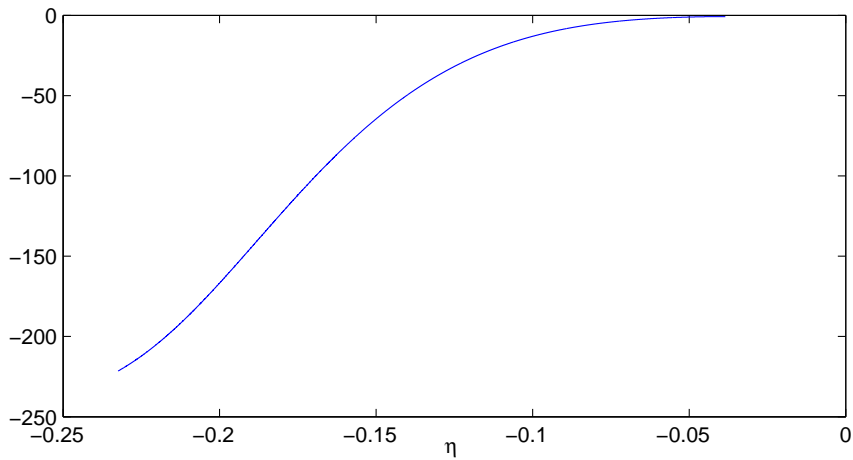}
  \caption{The potential $ \mathcal{V}_\mathcal{R}(\eta) $ vs. $\eta $ for
    new inflation with $ y = 1.322 $.
$ \mathcal{V}_\mathcal{R}(\eta) $ is {\bf attractive} during
    fast-roll and vanishes by the end of fast roll ($ \eta \sim -0.04 $).}
  \label{veta}
\end{figure}

We plot in fig. \ref{veta} $ \mathcal{V}_\mathcal{R}(\eta) $ vs. $ \eta $
for new inflation [eq.(\ref{wnue})] for the coupling $ y = 1.322 $ and a
total number of efolds equal to sixty.  We choose here the initial values
of $ \chi $ and $ \dot \chi $ such that their initial kinetic and potential
energies are equal. We see that the potential $
\mathcal{V}_\mathcal{R}(\eta) $ is {\bf attractive} in the fast-roll stage
and asymptotically vanishes by the end of fast roll $ \eta \sim -0.04 $.

\begin{figure}[h]
  \includegraphics[scale=0.8]{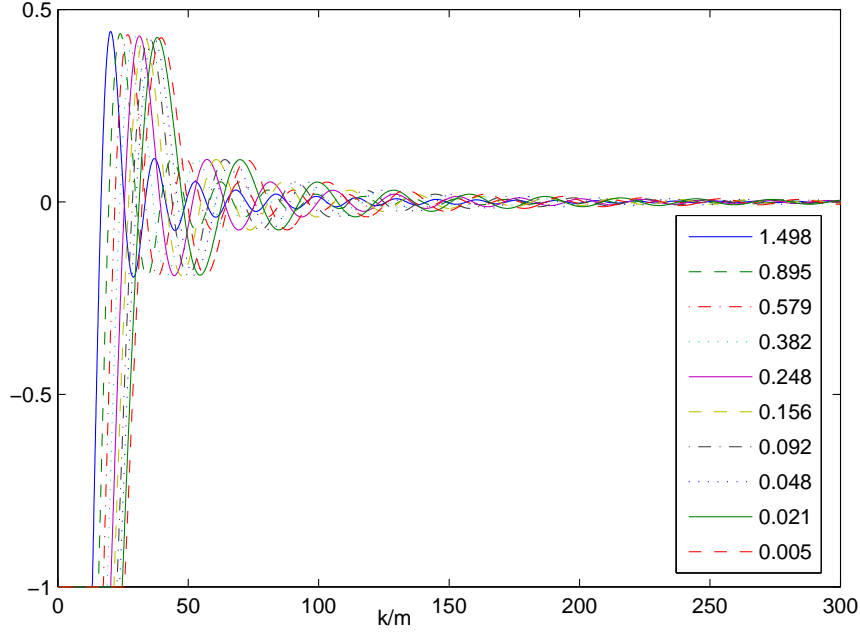}
  \caption{$ D_\mathcal{R}(k) $ vs. $ k/m $ for new inflation and ten
    different couplings $ 0.00536 < y < 1.498 $.
We see that the plots of $ D_\mathcal{R}(k) $
    for different couplings follow from each other by changing the scale in
    the variable $ k $ as summarized by eq.(\ref{Qk1}).}
  \label{dkflin4}
\end{figure}

We obtain the transfer function $ D_\mathcal{R}(k) $ by inserting $
\mathcal{V}_\mathcal{R}(\eta) $ into eq.(\ref{Dborn2}) and computing the
integral over $ \eta $ numerically. In fig.  \ref{dkflin4} we plot $
D_\mathcal{R}(k) $ vs. $ k/m $ for new inflation [eq.(\ref{wnue})] and ten
different couplings $ 0.00536 < y < 1.498 $ with a total number of efolds
equal to sixty. We see that $ D_\mathcal{R}(k) $ oscillates around zero and
therefore produces {\bf suppressions as well as enhancements} in the
primordial power spectrum [see eq.  (\ref{curvapot})].  $ D_\mathcal{R}(k)
$ vanishes asymptotically for large $ k $ as expected.

The first peak in $ D_\mathcal{R}(k) $ is clearly its dominant feature.
The $ k $ of this peak corresponds to $k$-modes which are today horizon
size and affect the lowest CMB multipoles (see below and table 2)
\cite{quadru1,quadru2}.

For small $ k $ the Born approximation to $ D_\mathcal{R}(k) $ yields large
negative values indicating that this approximation cannot be used in this
particular small $ k $ regime.  We introduce the scale $k_1$ by the
condition $ D_\mathcal{R}(k_1) = -1 $ and then just take $ D_\mathcal{R}(k) = -1 $ for
$ k \leq k_1 $. This corresponds to vanishing primordial power for the lowest
values of $ k $ [see fig. \ref{dkflin4}].

\medskip 

>From fig. \ref{dkflin4} we also see that the plots of $ D_\mathcal{R}(k) $
for different couplings follow from each other almost entirely by changing
the scale in the variable $ k $ as summarized by eq.(\ref{Qk1}).  Indeed,
the characteristic scale $ k_1 $ plays here a further important role.

\begin{figure}[h]
  \includegraphics[scale=0.8]{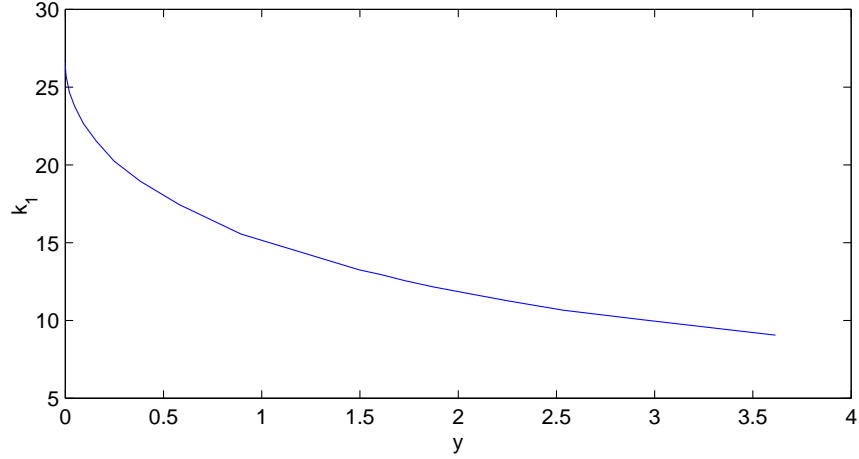}
  \caption{$ k_1 $ vs. $ y $ for new inflation.}
\label{k1y}
\end{figure}

Analysing $ \mathcal{V}_\mathcal{R}(\eta) $ and $ D_\mathcal{R}(k) $ for
different couplings $ y $ we find that they {\bf scale} with $ k_1 $.
Namely, 
\be\label{Qk1} 
   \mathcal{V}_\mathcal{R}(\eta) = k_1^2 \; Q(k_1 \;
   \eta) \quad , \quad D_\mathcal{R}(k) = \Psi\left(\frac{k}{k_1}\right) \; ,
\ee 
where $ Q(x) $ and $ \Psi(x) $ are universal functions. That is, $ Q(x) $
and $ \Psi(x) $ do not depend on the coupling $ y $ while $ k_1 = k_1(y) $.
We display $ k_1 $ vs. $ y $ in fig. \ref{k1y}.

\begin{figure}[h]
  \includegraphics[scale=0.8]{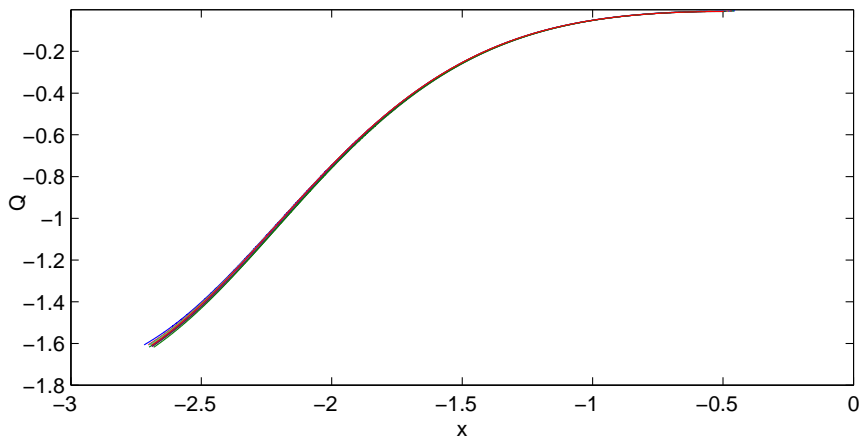}
  \caption{$ Q(x) $ for the ten values of $ y $ of fig.~\ref{dkflin4},
    according to eq.(\ref{Qx}).  All curves collapse to a common one
    proving the scaling properties eq. (\ref{Qk1}).}
  \label{Qxf}
\end{figure}

We obtain the function $ Q(x) $ from eq.(\ref{Qk1}) as,
\be\label{Qx}
Q(x) = \frac1{k_1^2} \; \mathcal{V}_\mathcal{R}\left(\frac{x}{k_1}\right)
\ee
We plot $ Q(x) $ in fig. \ref{Qxf} as follows from the r. h. s.
of eq.(\ref{Qx}) for ten different values of $ y $. We see that all the curves collapse
on a common curve proving the validity of the quasi-scaling properties eq. (\ref{Qk1}).

\section{MCMC analysis of CMB and LSS data including the early fast-roll
  inflationary stage}\label{mcmcsec}

In order to test the theoretical quadrupole depression predicted by fast-roll
inflationary stage against the current experimental data we performed a Monte
Carlo Markov Chains (MCMC) analysis of the commonly available CMB and LSS data
using the {\em CosmoMC} program \cite{lewis}.  

For LSS we considered SDSS (DR4). For CMB we first considered the three--years
WMAP data (with the second release of WMAP likelihood code) and small scale data
(ACBAR-2003, CBI2, BOOMERANG03). While this work was in progress the five--years
WMAP data were released, and we repeated our MCMC analysis almost completely
with these new data, using also the newer 2007 ACBAR release. Actually WMAP-3 or
WMAP-5 provide by far the dominant contribution and small scale experiments have
very little relevance for the quadruple depression issue.

In all our MCMC runs we have not marginalized over the SZ (Sunayev-Zel'dovich)
amplitude and have not included non-linear effects in the evolution of the
matter spectrum.  The relative corrections are in any case not significant
\cite{WMAP-3,mcmc}, especially in the present context.

{\em CosmoMC} is a publicly available open--source FORTRAN package that performs
MCMC analysis over the parameter space of the Standard Cosmological model and
variations thereof. The main observables in this approach are the correlations
among the CMB anisotropies and in particular: the TT (Temperature--Temperature),
the TE (Temperature--E\_modes), the EE (E\_modes--E\_modes) and the BB
(B\_modes--B\_modes) correlation multipoles (E\_modes and B\_modes are special
modes of the CMB polarization).  These multipoles can be numerically calculated
with very good accuracy, as functions of the cosmological parameters, from the
primordial power spectrum through programs such as CAMB (included in CosmoMC).
On the other side, experimental data provide a likelihood distribution for
multipoles, which is then turned into a likelihood for the cosmological
parameters through the MCMC method. We modified the {\em CosmoMC} code
introducing the transfer function $ D_\mathcal{R}(k) $ in the primordial power
spectrum according to eq.~(\ref{curvapot}).

We ran {\em CosmoMC} on pc clusters with Message Passing Interface (MPI),
producing from 10 to 24 parallel chains, with the `R-1' stopping criterion
(which looks at the fluctuations among parallel chains to decide when to
stop the run) set equal to 0.03. The statistical converge was also verified 
a posteriori with the help of the {\em getdist} program of {\em CosmoMC}.

\bigskip

The preferred reference model for slow--roll inflation cosmology is the
$\Lambda$CDM+$r $ model, that is the standard $\Lambda$CDM model, which has
six parameters\footnote{we use the standard ones of CosmoMC, that is the
  baryonic matter fraction $ \omega_b $, the dark matter fraction $\omega_c
  $, the optical depth $ \tau $, the ratio of the (approximate) sound
  horizon to the angular diameter distance $ \theta $, the primordial
  superhorizon power in the curvature perturbation at $ 0.05 ~{\rm
    Mpc}^{-1}, \; A_s $ and the corresponding scalar spectral index $ n_s
  $}, augmented by the tensor-scalar ratio $ r $. Indeed, the current
experimental accuracy provides sensible bounds only for the first order
parameters $\epsilon_v$ and $\eta_v$, through their standard relation to
the scalar spectral index $ n_s $ and the ratio $ r : \; n_s-1=2\,\eta_v-
6\,\epsilon_v , \;  r=16\,\epsilon_v $.  Specific slow--roll scenarios, such as
those based on new (small--field) or chaotic (large--field) inflation,
connect in a model--dependent way $ \epsilon_v $ and $ \eta_v $ to free
parameters in the inflaton potential and thus typically lead to specific
theoretical constraints in the $ (n_s,\,r) $ plane \cite{mcmc}.

\medskip

We point out that we used the default {\em CosmoMC} pivot scale $ k_0 = 0.05 ~ {\rm
  Mpc}^{-1}$ rather than the customary WMAP choice of $k_0 = 0.002\,{\rm
  Mpc}^{-1}$. As evident from eq.~(\ref{potBD}) this leads to a small difference
with respect to the WMAP choice in the definition itself of the tensor-scalar ratio
$r$. In particular, the CosmoMC $r$ is roughly 10\% larger than the WMAP one.

\subsection{MCMC analysis without quadrupole suppression: $ D_\mathcal{R}(k) = 0$.}

Let us present our MCMC analysis with the standard slow-roll primordial power
eq.(\ref{potBD}). That is, without including the early fast roll stage and
therefore vanishing transfer function $ D_\mathcal{R}(k) $.

\begin{figure}[h]
\includegraphics[height=10cm]{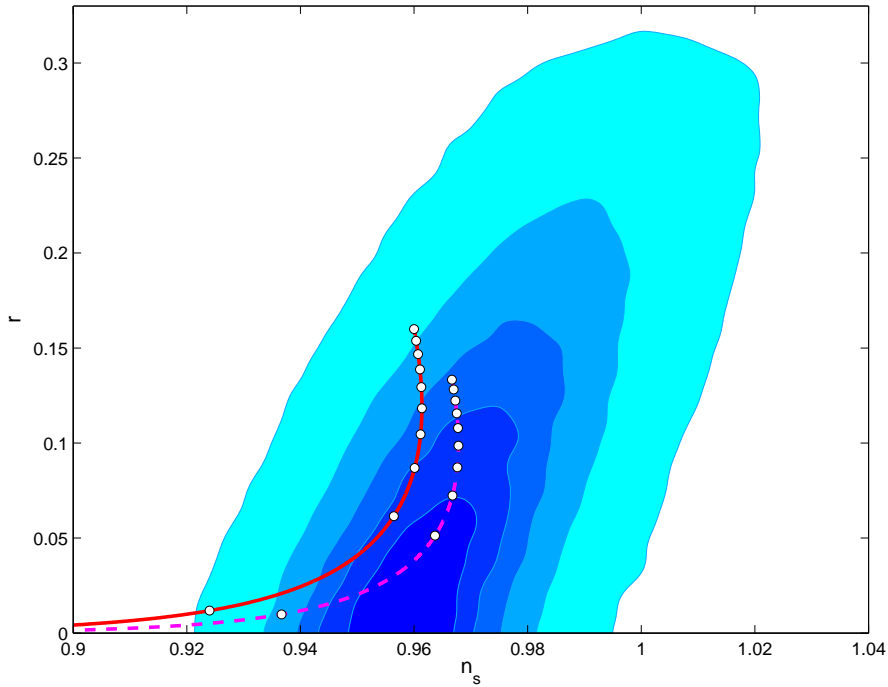}
\caption{Binomial New Inflation compared to $\Lambda$CDM+$r $ model in the
  $(n_s,\,r)$ plane. The color--filled areas correspond to $ 12\%, \; 27\%,
  \; 45\%, \; 68 \% $ and $ 95 \% $ confidence levels for $\Lambda$CDM+$r$
  according to WMAP-3, small scale CMB and SDSS data.  $C_\mathrm{BNI}$ is
  the solid red curve for $ N=50 $ or the dashed magenta curve for $N=60$. The
  white dots corresponds to the values $ 0.01 + 0.11*n , \; n=0,1, \ldots,  9 $, of the
  variable $ z $ in eq.~\eqref{BNI}, starting from the leftmost ones.
  The quartic coupling $ y $ instead increases monotonically starting from the
  uppermost dots, which corresponds to the free-field, purely quadratic
  inflaton potential $ y = 0 $. We see that very small values of $ r $ {\bf are excluded}
since they correspond to $ n_s < 0.92 $ outside the $ 95 \% $ confidence level
contour.}
\label{nsr}
\end{figure}

For instance, in the simplest binomial realization of new inflation described by
the inflation potential of eq.~\eqref{nueva} or eq.~\eqref{wnue}, $ n_s $ and $
r $ are constrained to the curve $ C_\mathrm{BNI} $ (BNI stands
for {\em Binomial New Inflation}) parametrized by the quartic coupling $ y $ as
\cite{mcmc}:
\begin{equation}\label{BNI}
  n_s=1- \frac{y}{N}\, \frac{3 \,z + 1}{(1-z)^2}  \;,\qquad
  r=\frac{16 \; y}{N} \frac{z}{(1-z)^2} \;, \qquad
  y=z-1-\log z \;,\qquad z = \frac{y}{8} \; \chi^2  \;, \qquad 0<z<1 \;.
\end{equation}
This situation is clearly displayed
 in fig.~\ref{nsr} in which the curve $
C_\mathrm{BNI} $, for the two choices $ N=50 $ and $ N=60 $, is drawn over the
contour plot of the likelihood distribution for $ n_s $ and $ r $ in the $
\Lambda$CDM+$r $ model obtained with CosmoMC, using the WMAP-3, small-scale CMB
and SDSS data. Practically, the same contour plot applies when WMAP-5 and
ACBAR-2007 are used.

The likelihood $ L $, as function of the whole set of parameters, provides a
quantitative measure of the power of a given model to fit the multipoles $
C_{\ell}^\gamma $. As customary, we set $ -2\log L = \chi_L^2 $, although it is
well known that, due particularly to cosmic variance, the shape of $ L $, as
function of the $ C_{\ell}^\gamma $, is not Gaussian especially for low $ \ell$.

\medskip

Now, as evident from eq.~(\ref{BNI}) and fig.~\ref{nsr}, one could expect
from  the $ \Lambda\mathrm{CDM} $ model constrained to $ C_\mathrm{BNI} $
 a fit to the data not as good as in the $\Lambda$CDM+$r $ model 
since the
current data seem to favor smaller values for $ r $. Indeed we find
\begin{equation}\label{dchi21}
  {\rm min}\,\chi_L^2(\Lambda\mathrm{CDM}+r~\mathrm{on}~C_\mathrm{BNI}) 
  -{\rm min}\,\chi_L^2(\Lambda\mathrm{CDM}+r) \simeq 0.4 \;.
\end{equation}
This result was obtained for $ N=50 $ by direct minimization of $ \chi_L^2 $ in
the neighbourhood of $ C_\mathrm{BNI} $, using the data of a large collection of
long chain runs (with a grandtotal of almost two million steps) for the $
\Lambda$CDM+$r $ model with the WMAP-3, small--scale CMB and LSS data. The flat
priors on the cosmological parameters were the standard ones of CosmoMC, that is
\begin{center}
  \begin{tabular}{l l l}
    $0.005<\omega_b<0.1 \quad ,$ & $\;0.01<\omega_c<0.99 \quad ,$ &$\,0.5<\theta <10$ \\
    $0.01<\tau<0.8 \quad , $ & $\;2.7 < \log(10^{10} A_s)<4 \quad , $ & $\;0.5<n_s<1.5$ \\
  \end{tabular} 
\end{center}
while for the tensor-scalar ratio we imposed as prior
\begin{equation*}
  0 < r < 0.35 \;.
\end{equation*}
We repeated the same analysis with the WMAP-5, ACBAR-2007 and SDSS data,
this time setting $N=60$, with no statistically significant change.

\medskip

Another approach, that unlike the direct minimization of $ \chi_L^2 $ over $
C_\mathrm{BNI} $ does take advantage of the explicit analytic parametrizations
in eq.~\eqref{BNI}, is to use the single variable $ z $ as MCMC parameter,
instead of the constrained pair $ (n_s\,r) $, with a flat prior over all the
allowed range $ 0<z<1 $. Let us call $\Lambda$CDM$z|C_\mathrm{BNI}$ the
$6$-parameter model $ \Lambda {\rm CDM} $ constrained on $C_{\rm BNI}$ using the
variable $z$. Then, we find that taking into account the natural fluctuations
due to the large number of data (which make the likelihood landscape over the
MCMC parametrs quite complex) and the various approximations and numerical
errors in the theoretical calculation of the multipoles, the increasing in $
\chi_L^2 $ due to the $C_{\rm BNI}$ eq.~\eqref{BNI} constraint compared to the $
\Lambda {\rm CDM} $ model essentially vanishes [see Table I below].
  
\medskip

For completeness and reference, we report in Table I our best fit (or most likely)
values for the MCMC cosmological parameters, as well as the absolute value
of our best likelihoods, which of course depend on the specific datasets
used, that is WMAP-3, small scale CMB and SDSS. We report in the first line
of Table I our best fit
for the standard $\Lambda$CDM model, which has six free parameters since $ r $
is set to zero by fiat.
\begin{table}\label{T1}
  \begin{tabular}{c|c|c|c|c|c|c|c|c|}
    & $~~10\,\omega_b~~$ & $~~~~\omega_c~~~~$ & $~~10\,\theta~~$ & $~~~~\tau~~~~$ 
    & $~~10^{9}A_s~~$ & $~~~~n_s~~~~$ & $~~~~r~~~~$ & $~~~\chi_L^2/2~~~$ \\ \hline
    $\Lambda$CDM
    & 0.224 & 0.106 & 1.041 & 0.886 & 2.072 & 0.959 & 0 & 2713.906 \\ \hline
    $\Lambda$CDM$+r$ 
    & 0.224 & 0.107 & 1.042 & 0.831 & 2.054 & 0.960 & 0.009 & 2713.972\\\hline
    $\Lambda$CDM$+r$ on $C_\mathrm{BNI}$ 
    & 0.223 & 0.106 & 1.040 & 0.848 & 2.047 & 0.956 & 0.059 & 2714.166 \\\hline
    $\Lambda$CDM$z|C_\mathrm{BNI}$ 
    & 0.222 & 0.107 & 1.041 & 0.877 & 2.065 & 0.958 & 0.069 & 2713.918 \\\hline
  \end{tabular} 
  \caption{Best fit values for the MCMC cosmological parameters without
    quadrupole suppression, using WMAP-3, small--scale CMB and SDDS. 
    $ C_\mathrm{BNI}$ means the curve on which $n_s$ and
    $r$ are constrained in Binomial New Inflation (BNI), eq.~\eqref{BNI} with $N=50$.
    $\Lambda$CDM+$r$ on $ C_\mathrm {BNI} $ means the $\Lambda$CDM+$r$ 
    model constrained on $C_\mathrm{BNI}$. $\Lambda$CDM$z|C_\mathrm{BNI}$
    denotes the $\Lambda$CDM model constrained on $ C_\mathrm{BNI} $ using the 
    single variable $z$ eq.~\eqref{BNI} as MCMC variable
    instead of the constrained pair $(n_s,r)$.}
\end{table}

\medskip

It should be noted that the likelihoods difference between $\Lambda$CDM+$r$ and
its direct restriction to $ C_\mathrm{BNI} $, (that is $ \Lambda$CDM +$r$ on
$C_\mathrm {BNI}$), is mostly due to the SDSS data, which together with the
CMB data place an upper bound on $ r $ twice more stringent than WMAP-3
alone.  Indeed, when only the WMAP-3 data were used, we verified that no
significant likelihoods difference was exhibited. This situation changes
slightly when WMAP-5 is used, with $\chi_L^2$ increasing approximately by 0.2 from
$\Lambda$CDM+$r$ to $\Lambda$CDM$+r$ on $C_\mathrm{BNI}$, since WMAP-5 alone
puts a tighter bound on $r$ than WMAP-3 alone ($0.43$ vs. $0.65$ at 95\% CL
\cite{WMAP-5}).

\medskip

It is evident that, as far as most likely values of the cosmological parameters
are concerned, the fit with the constraint $ C_\mathrm{BNI} $ included, either
with or without $ z $ as MCMC parameter, does not determine any statistically
significant change, except of course for $ n_s $ and $ r $ themselves. In
particular, with respect to the $ \Lambda \mathrm{CDM}+r $ results, the most
likely value of $ n_s $ is practically unchanged, while that of $r$ changes from
values of order $ 10^{-2} $ (or just $0$ in $\Lambda$CDM) to values such as $
0.059 $ and $ 0.069 $ (see Table I).

\medskip

Concerning marginalized distributions, we find no significant changes but for $
n_s $ and $ r $. These results are very close to those in ref.  \cite{mcmc},
where {\em trinomial new inflation}, with a possibly asymmetric potential, was
considered. In particular, the marginalized distribution for $ r $ shows a broad
but clear peak centered near the most likely value as in \cite{mcmc}. In the
present context of {\em binomial} new inflation, we find that
$r=0.089_{-0.05}^{+0.044}$ with $r>0.023$ at 95\% CL.

\medskip 

All together, these results show that, as far as pure data fitting is concerned,
the 6-parameter $\Lambda\mathrm{CDM} z|C_\mathrm{BNI}$ model is just as good as
the standard 6-parameter $\Lambda$CDM model. More generally speaking, we may say
that current CMB and LSS data together, {\bf without} any theoretical
constraint, put only an upper bound on $ r $ (namely $ r < 0.20 $ with 95\%
confidence level in the most recent WMAP-5 analysis \cite{WMAP-5}). Therefore,
any inflation--based 6-parameter model (as the $\Lambda$CDM$z|C_\mathrm{BNI}$
model, for instance) predicting a value of $ r $ well below 0.2 is as likely as
the $\Lambda$CDM model itself.  This means that the theoretical grounds of a
given model take a more important role in the analysis and interpretation of the
CMB and LSS data. For instance, from an inflationary viewpoint, the choice that
$ r $ vanishes exactly appears unlikely and unphysical. Notice that $ n_s - 1 =
r = 0 $ corresponds to a singular and critical (massless) limit where the
inflaton potential vanishes \cite{mcmc}, while The MCMC analysis for both the binomial
and trinomial new inflationary models yield {\bf lower bounds} for $ r $.

\medskip

In order to asses the statistical relevance of the
quadrupole suppression, we determine in the best fit $\Lambda$CDM model, 
the probability that there is at least one multipole, regardless of 
$\ell$, smaller that 20\% of the theoretical mean value.
We obtained 0.06126 for such probability. 
Thus, in the $\Lambda$CDM, the observed quadrupole realizes a rather 
unlikely event which has only a 6\% probability.
Therefore, it makes sense to search for a cosmological explanation
of the quadrupole depression beyond the $\Lambda$CDM model.

\subsection{MCMC analysis including the quadrupole suppression: $ D_\mathcal{R}(k) \neq 0 $.}

Let us now further develop this argument by considering the quadruple
depression, avoiding the a priori dismissal based on the simple invocation
of cosmic variance or experimental inaccuracy. In the standard $\Lambda$CDM
model the simplest, purely phenomenological way to decrease the low
multipoles is to introduce a infrared sharp cut in the primordial power
spectrum of the curvature fluctuations. That is, one assumes that
$P_\mathcal{R}(k)=0$ for $k<k_1$ and treats $k_1$ as a new MCMC parameter
to be fitted against the data.  It is actually not necessary to include
also a cut on the tensor power spectrum, since it would lead to changes
certainly not appreciable within the current experimental accuracy.

With this procedure we obtained, using either the WMAP-3 data alone or both
CMB and LSS data:
 \begin{equation*}
   {\rm min}\,\chi_L^2(\Lambda\mathrm{CDM}+{\rm sharpcut}) 
   - {\rm min}\,\chi_L^2(\Lambda\mathrm{CDM}) \simeq -1.4 
\end{equation*} 
 
\begin{figure}[ht]
\includegraphics[height=8cm]{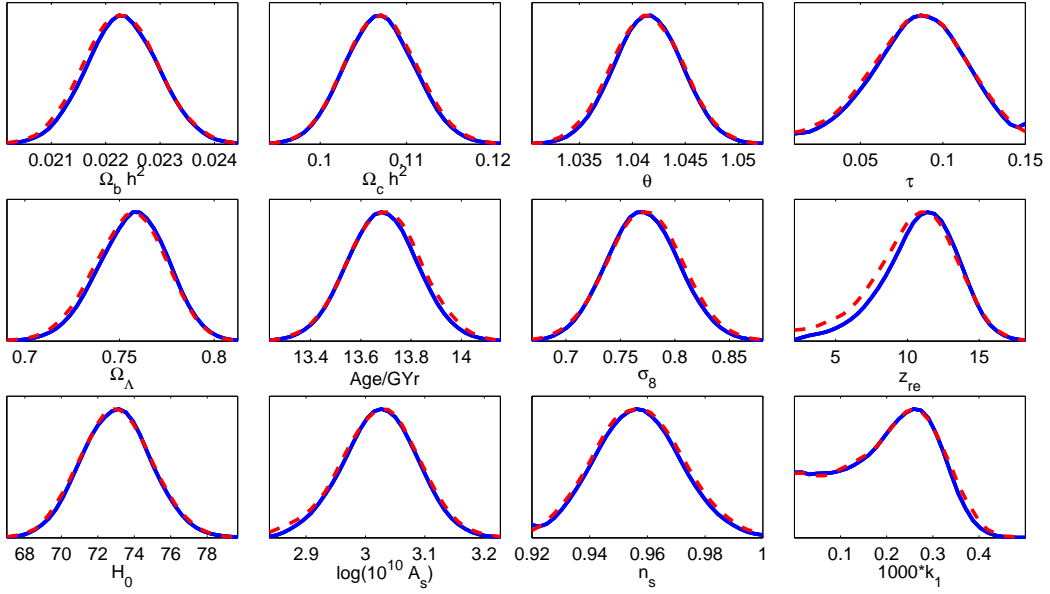}
\caption{Marginalized distributions (solid blue lines) and mean likelihoods
  (red dotted lines) for the parameters of the
  $\Lambda\mathrm{CDM}+$sharpcut model.}
\label{sharpcut}
\end{figure}

This result is slightly better than the one reported in ref.\cite{WMAP-3},
but still the likelihood gain hardly compensates the price of a new
parameter, especially because its nature appears quite {\em ad hoc}. In
fig.~\ref{sharpcut} we plot the marginalized probabilities and mean
likelihoods of the seven MCMC parameters plus other standard derived
parameters in the CMB+LSS case. In the WMAP-3 alone case these plots are almost
identical. There are no significant changes from $\Lambda\mathrm{CDM}$ to
$\Lambda\mathrm{CDM}+$sharpcut in their common parameters, in either most
likely values or marginalized distributions. The distribution of the new
cutoff parameter $ k_1 $ shows a well defined peak centered on its most
likely value (ML), which corresponds to today's physical wavelength
 \begin{equation*}
   (k_1)_{\rm ML} = \begin{cases} 
     0.291 \; ({\rm Gpc})^{-1} \quad \textrm{(WMAP-3 only)}\\
     0.272 \; ({\rm Gpc})^{-1} \quad \textrm{(CMB+LSS)} 
   \end{cases}  \qquad   (\Lambda{\rm CDM} + {\rm sharpcut}) \;,
 \end{equation*}
that is of the order of today's inverse Hubble radius, as expected.

\medskip

\begin{figure}[ht]
\includegraphics[height=8cm]{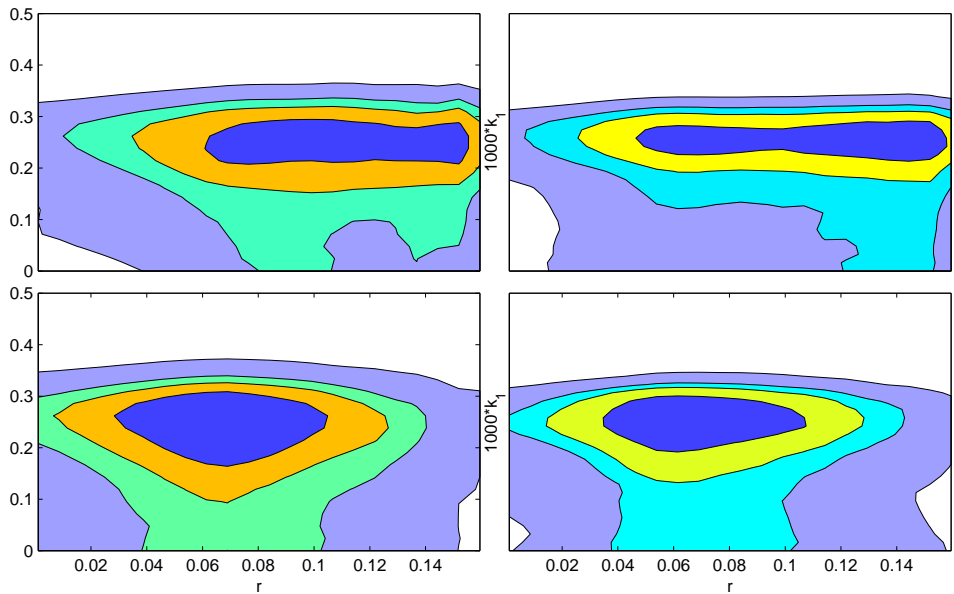}
\caption{Marginalized pair distributions in the $(r,\,k_1)$ plane, at 20\%,
  41\%, 68\% and 95\% CL, with the WMAP-3 data alone in the top panels and with
  the CMB+SDSS data in the bottom panels.  The left panels refer to the
  $\Lambda$CDMz$|C_\mathrm{BNI}+$sharpcut model, while the right ones are for
  the $\Lambda$CDM$z|C_\mathrm{BNI}+$fastroll model.  $ k_1 $ is in $ {\rm
    Mpc}^{-1} $.}
\label{pdrk1}
\end{figure}

Introducing the infrared sharp cut on $ P_\mathcal{R}(k) $ in the
$\Lambda\mathrm{CDM}z|C_\mathrm{BNI}$ model we find sizably different gains
\begin{equation*}
  {\rm min}\,\chi_L^2(\Lambda\mathrm{CDM}z|C_\mathrm{BNI}+{\rm sharpcut}) 
  - {\rm min}\,\chi_L^2(\Lambda\mathrm{CDM}z|C_\mathrm{BNI}) =
  \begin{cases}
    -1.4 \quad \textrm{(WMAP-3 only)} \\
    -0.8 \quad \textrm{(CMB+LSS)}
   \end{cases}
\end{equation*}
As before, the difference is due to the tighter bound on $ r $ due to the
inclusion of the SDSS data. In fact, the most likely values (ML) of $ k_1 $ and $ r $
corresponding to $ {\rm min}\,\chi_L^2(\Lambda\mathrm{CDM}z|C_\mathrm{BNI}+{\rm
  sharpcut}) $ are given in Table II.
\begin{table}
  \begin{tabular}{c|c|c|c|}
    $\Lambda$CDM$z|C_\mathrm{BNI}+$sharpcut & $~~k_1$ (best fit)~~ &  
    $~~~~r$ (best fit)~~ &  $~~~r$ (95\% CL)~~\\ \hline
    WMAP-3 only &  0.275  $ {\rm Gpc}^{-1}$ & 0.150 & $> 0.023$ \\ \hline
    CMB+LSS  &  0.268 $ {\rm Gpc}^{-1}$ & 0.051 & $> 0.018$ \\ \hline
  \end{tabular} 
  \caption{The most likely values of $ k_1 $ and $ r $ and the lower bound on
    $r$ in the $\Lambda\mathrm{CDM}z|C_\mathrm{BNI}+{\rm sharpcut})$ model.}
\end{table}
The marginalized probabilities in the $ (r,\,k_1) $ plane (converting to a
flat prior on $ r $) are shown in the two left panels of fig.~\ref{pdrk1}.
There are no significant changes on the other cosmological parameters. 

\medskip

This situation is also reproduced when the fastroll stage is included, that is
when the fast roll transfer function $ D_\mathcal{R}(k) $ eq. (\ref{Dborn2}) and
fig. \ref{dkflin4} is used, treating the scale $ k_1 $ in eq. (\ref{Qk1}) as a
MCMC parameter.

That is, in the MCMC analysis with fast-roll included, we use the initial power
spectrum eq.(\ref{curvapot}) which is modified by the fast-roll transfer
function $ D_\mathcal{R}(k) $.  We computed once and forever $ D_\mathcal{R}(k)
$ from eq. (\ref{Dborn2}) [see fig. \ref{dkflin4}].  $ D_\mathcal{R}(k) $ is a
function of $ k $ and $ k_1 $ with the scaling form eq.(\ref{Qk1}), $ \Psi(x) $
being an universal function.

We then find
\begin{equation*}
  {\rm min}\,\chi_L^2(\Lambda\mathrm{CDM}z|C_\mathrm{BNI}+{\rm fastroll}) 
  - {\rm min}\,\chi_L^2(\Lambda\mathrm{CDM}z|C_\mathrm{BNI}) =
  \begin{cases}
    -1.8 \quad \textrm{(WMAP-3 only)} \\
    -1.2 \quad \textrm{(CMB+LSS)}
   \end{cases} \;
\end{equation*}
Correspondingly, in Table III we report the most likely values (ML) of $k_1$ and $r$  
(we report also the best fit for the quartic coupling $ y $ for future use)
\begin{table}
  \begin{tabular}{c|c|c|c|c|}
    $\Lambda$CDM$z|C_\mathrm{BNI}+$fastroll& $~~k_1$ (best fit)~~
    & $~~r$ (best fit)~~ & $~~~r$ (95\% CL)~~ & $~~~y$ (best fit)~~  \\ \hline
    WMAP-3 only &  0.249 $ {\rm Gpc}^{-1} $  & 0.146 &$>0.018$ & 0.031\\ \hline
    CMB+LSS  &  0.266  $ {\rm Gpc}^{-1} $ & 0.058 & $>0.018$ & 1.322 \\ \hline
  \end{tabular} 
  \caption{The most likely values of $ k_1 $, $ r $ and the quartic coupling $ y
    $ and the lower bound on $r$ in the 
    $\Lambda\mathrm{CDM}z|C_\mathrm{BNI}+{\rm fastroll})$ model.}
\end{table}
where we used the marginalized probability in the $ (r,\,k_1) $ plane as shown
in the two right panels of fig.~\ref{pdrk1}. Here
$\Lambda\mathrm{CDM}z|C_\mathrm{BNI}+{\rm fastroll}$ denotes the
$\Lambda\mathrm{CDM}z|C_\mathrm{BNI}$ model with the fast roll
$D_\mathcal{R}(k)$ included.

\medskip

We see a clear peak in $ y $ {\bf when} fast-roll or a sharp cut are introduced 
in the CMB$+$SDSS fits.

\begin{figure}
\includegraphics[height=10cm]{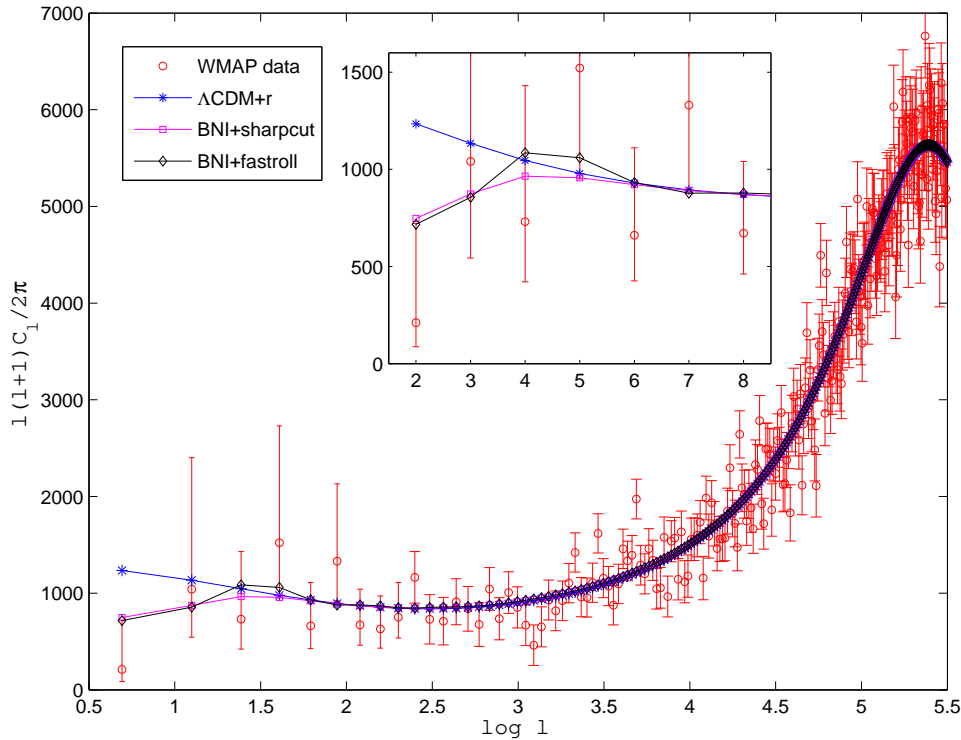}
\caption{Comparison, with the experimental WMAP-3 data, of the theoretical
  $ C_{\ell}^{\rm TT} $ multipoles computed in the best fit point of the
  various models of the main text. The error bars in the plotted range of
  $ \ell $ are mostly due to cosmic variance. The insert contains an
enlargement in linear scale of the first seven multipoles. 
BNI stands for binomial new inflation. 
The  $ C_{\ell}^{\rm TT} $ units are $ [\mu \; K^2] $ and they are plotted
as functions of the natural logarithm of $ \ell $.
Error bars of the WMAP-3 data are one-sigma ($68\%$c.l.).}
\label{clsfig}
\end{figure}

We see that the gains in likelihood are {\em more significant in the
fast-roll case} than in the sharpcut case. Clearly, this fit improvement
through power modification by fastroll over power reduction by sharpcut is
too small to constitue a real experimental evidence.  But still, it is very
interesting that the theoretically well founded approach based on fastroll
works {\bf better} than the purely phenomenological cutoff.  This may be
appreciated also from fig.~\ref{clsfig} where the best fit for the $ C_{\ell}^{\rm
  TT} $ multipoles are compared to the experimental data at low $ \ell $. We see
that the oscillatory form of the fastroll transfer function
$ D_\mathcal{R}(k) $, by {\bf depressing as well as enhancing} the primordial
power spectrum at long wavelengths, leads also to new superimposed {\bf oscillatory corrections} 
on the multipoles. As far as fitting to current data is concerned, such
corrections are more effective than the pure reduction caused by a
sharp cutoff.

\medskip

\begin{figure}
\includegraphics[width=14cm,height=10cm]{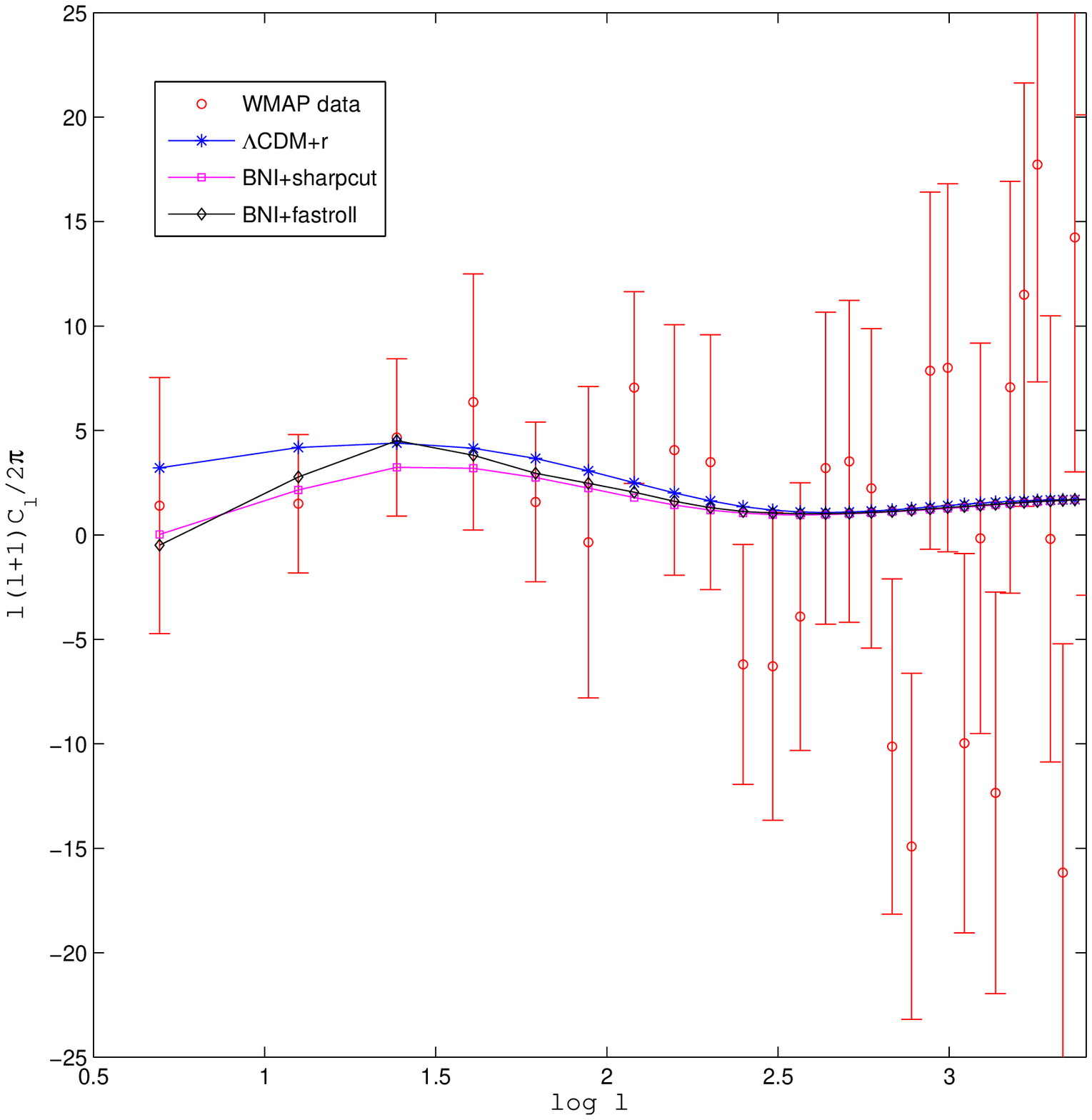}
\caption{Comparison, with the experimental WMAP-3 data, of the theoretical $
  C_{\ell}^{\rm TE} $ multipoles computed in the best fit point of the
  $\Lambda$CDM+$r $ model, fast-roll and sharpcut models. Notice that for $
  C_2^{\rm TE} $ and $ C_3^{\rm TE} $ fast-roll and sharpcut models yield rather
  similar results (and better than the $\Lambda$CDM+$r$ model), while for $ \ell
  = 4 $ fast-roll produces a value closer to WMAP-3 than sharpcut. 
The  $ C_{\ell}^{\rm TE} $ units are $ [\mu \; K^2] $ and they are plotted
as functions of the natural logarithm of $ \ell $.
Error bars of   the WMAP-3 data are one-sigma ($68\%$c.l.).}
\label{clsTE}
\end{figure}

We did not display in figs. \ref{clsfig}  and \ref{clsTE} 
the  $\Lambda$CDM$+$sharpcut results since they are indistinguishable 
from the BNI$+$sharpcut values.

We plot in fig. \ref{clsTE} the best fit for the $ C_{\ell}^{\rm TE} $
multipoles compared to the experimental data at low $ \ell $.  We see that for $
\ell = 2 $ and $ \ell = 3 $ fast-roll and sharpcut models yield rather similar
results (and better than the $\Lambda$CDM$+r$ model) while for $ \ell = 4 $
fast-roll produces a value closer to WMAP-3 data than sharpcut.

\begin{figure}
\includegraphics[width=14cm,height=10cm]{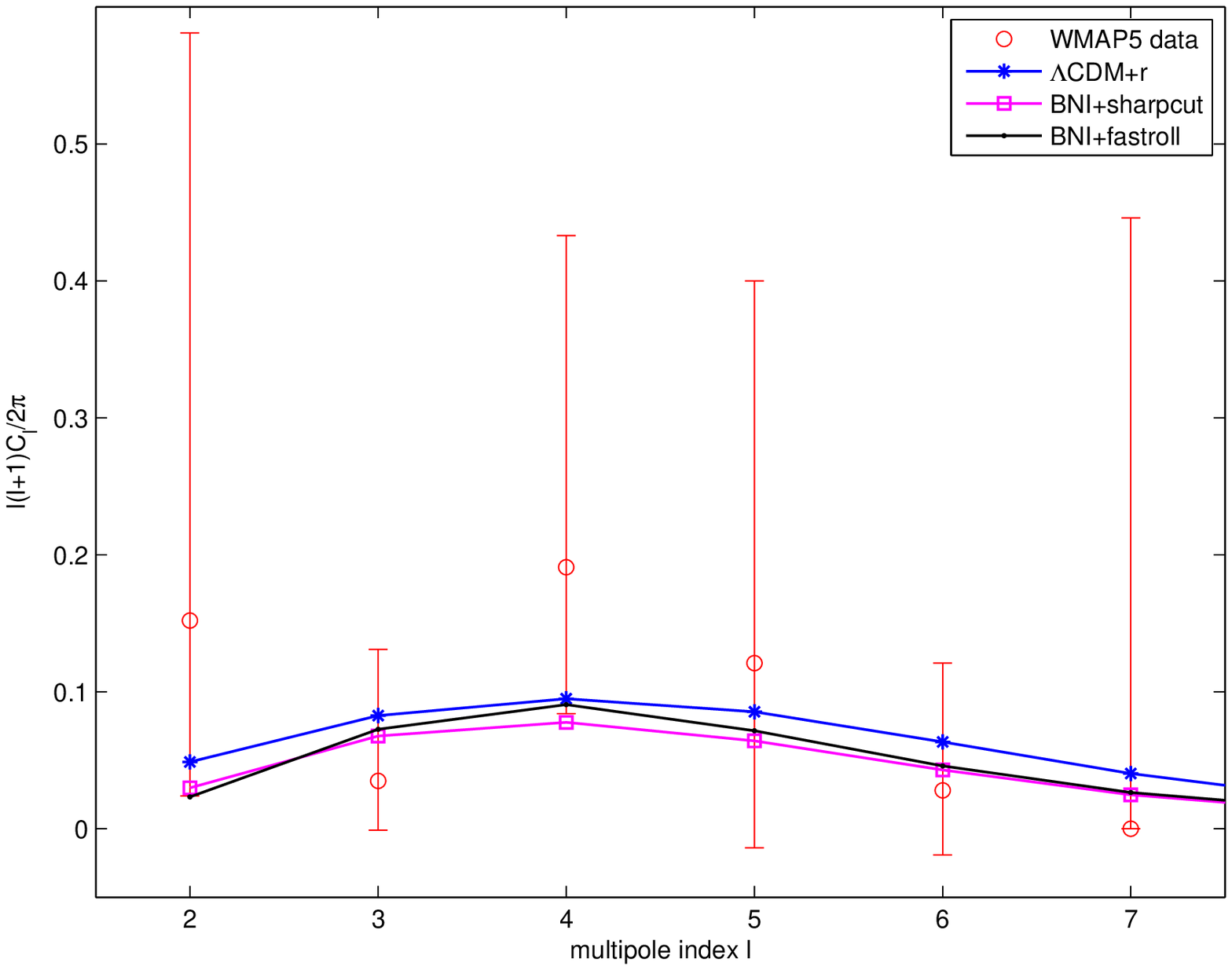}
\caption{Comparison, with the experimental WMAP-5 data, of the 
theoretical $ C_{\ell}^{\rm EE} $ multipoles computed in the best fit  
point to the WMAP-5 data of the $\Lambda$CDM+$r $ model, fast-roll 
and sharpcut models as functions of $ \ell $.
Error bars of the WMAP-5 data are one-sigma ($68\%$c.l.).
Notice that both fast-roll and sharpcut models produce
a depression of the low EE multipoles including the EE quadrupole.
The  $ C_{\ell}^{\rm EE} $ units are $ [\mu \; K^2] $.}
\label{clsEE}
\end{figure}

We plot in fig. \ref{clsEE}  the $ C_{\ell}^{\rm EE} $ multipoles computed
in the best fit point to the WMAP-5 data 
compared to the experimental WMAP-5 data at low $ \ell $. 
We see that both fast-roll and sharpcut models produce a reduction
of the low EE multipoles including the EE quadrupole.

\medskip

Our fits imposing $ C_\mathrm {BNI} $ predict {\bf non-zero} lower bounds on $ r
$: at 95\% CL, we find that $r>0.023$ when no cutoff is 
introduced, while $r>0.018$ when either the sharp cutoff or the fastroll $D(k)$ are introduced.  
The best
fit values of the other cosmological parameters remain practically unchanged as
compared to $\Lambda$CDM. Similarly their marginalized probability 
distributions
are almost unchanged, with the natural exception of $n_s$, which in BNI has a
theoretical upper limit [see eq.~(\ref{cotns})].

\subsection{Real Space Two Point TT-Correlator}

\begin{figure}
\includegraphics[height=10cm]{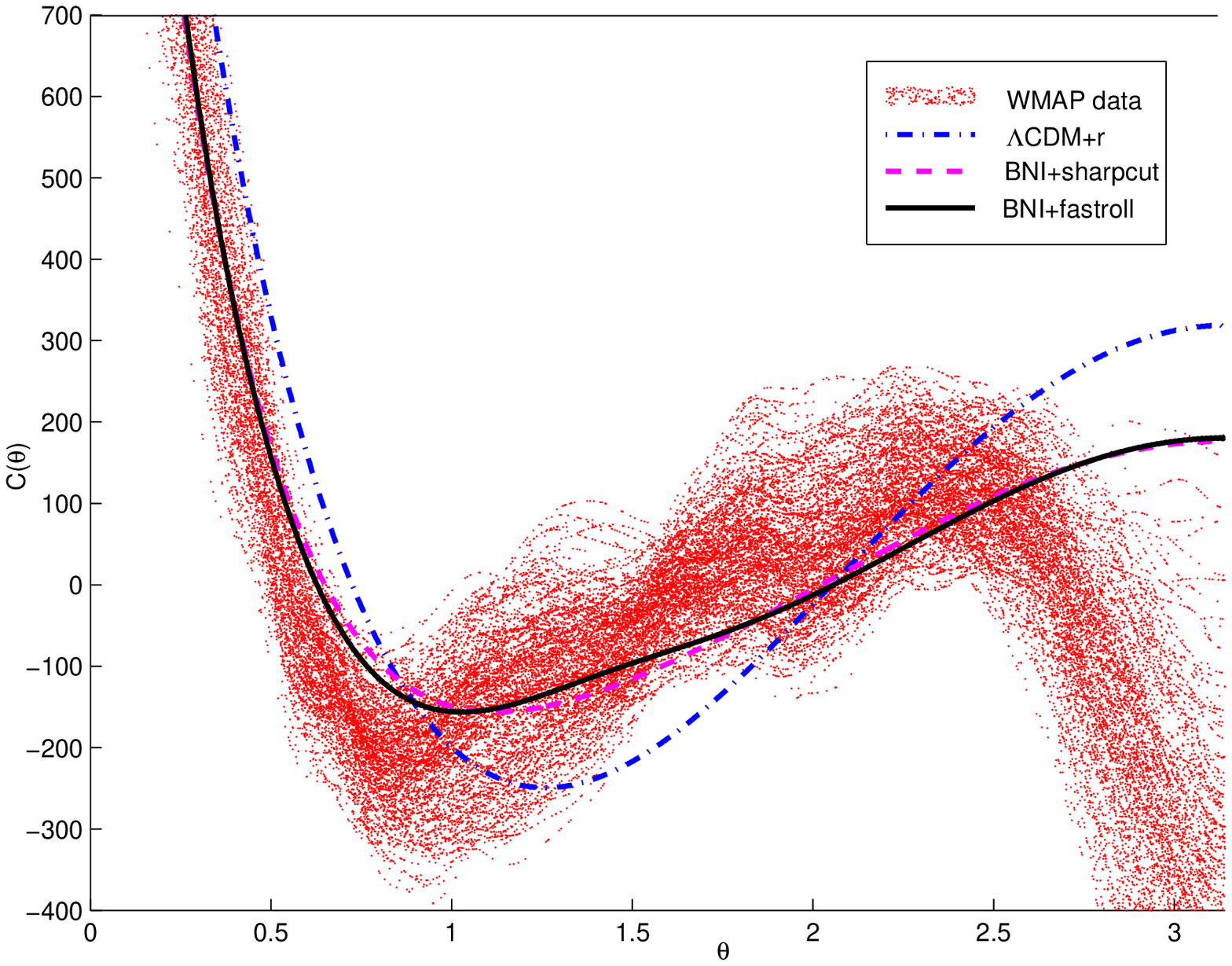}
\caption{The real space two point TT correlation function 
$ C^{TT}(\theta) $ for
  $\Lambda$CDM, sharpcut and fast-roll models vs. the angle $ \theta $.  The
  $\Lambda$CDM correlator differs from the two others only for large angles $
  \theta \gtrsim 1 $.  Since all $l$-modes besides the lowest ones are
  practically identical in the three cases, this shows how important are the
low multipoles in the large angle correlations. Also shown are the WMAP data. 
The truly observed correlator runs approximately in the middle of the red band.
The width of the data band is mostly due to the cosmic variance.
The WMAP $ C^{TT}(\theta) $ plotted here may not coincide, especially for
the largest values $ \theta \sim \pi $, with the correlator directly 
measured from sky maps due to the pixel weighting in the WMAP data 
analysis. The $ C^{TT}(\theta) $ units are $ [\mu \; K^2] $.}
\label{corrR}
\end{figure}

We display in fig. \ref{corrR} the real space two point TT-correlation function
$ C^{TT}(\theta) $ for $\Lambda$CDM, sharpcut and fast-roll models,
$$
C^{TT}(\theta) = \frac1{4 \, \pi} \sum_{l=2}^{\infty}(2\,l+1) \; C_l^{TT} \; 
P_l(\cos \theta) \; .
$$
We see that the $\Lambda$CDM correlator
becomes really different from the two others only for large angles $ \theta \gtrsim 1 $.
Since all $l$-modes besides the lowest ones are practically identical in the three cases, 
this shows how dominant  are the low multipoles in the large angle correlations. 
We also show the WMAP data, the width of the data is mostly due to the cosmic variance.

As is clear from fig. \ref{corrR}, both fast-roll and sharpcut models reproduce the
two point correlator  $ C^{TT}(\theta) $ better than the pure slow-roll  $\Lambda$CDM$+r$
model.

\section{ The Total Number of e-folds of Inflation $ N_{tot} = N + 6 \sim 66 $}

Another interesting observation is possible concerning the number $ N $ of
efolds since horizon exit till the end of inflation. First of all let us
clarify why in all our MCMC runs we keep $ N $ fixed. The reason is that the
main physics that determines the value of $ N $ is {\bf not} contained in
the available data but involves the reheating era. Therefore, although
technically possible, it is not reliable to fit $ N $ solely with the CMB
and LSS data within a pure, near scale-invariant slow roll scenario. On the
other hand, the quadruple depression allows to set an absolute wavelength
scale for the primordial power, so we can check the consistency of our
assumptions about $ N $ which fixes the total number of efolds of inflation.

In the case of the $\Lambda$CDM$z|C_\mathrm{BNI}+$fastroll model with the 
CMB+LSS datasets, the most likely value of the quartic coupling $ y $ is 
slightly larger than unity. Then from fig.~\ref{k1y} we read a value $
\sim 14 $ for the ratio $ k_1/m $ at horizon exit. 

It is important to compare the quadrupole mode scale $ k_Q $ with the scale $ k_1 $
that characterizes the fast-roll stage.

The physical quadrupole ($ l = 2 $) wavemodes today $ k_Q $ are related to
the particle horizon today $ \eta_0 $ by
$$
 k_Q  \; \eta_0 = 3.342\ldots \; ,
$$
where the spherical Bessel function $ j_2(k \;\eta_0) $ takes its maximun value, 
and $ \eta_0 $ is given by \cite{quadru1}
$$
\eta_0 = \frac{3.29}{H_0}
$$
when one takes into account the acceleration of the universe expansion
for $ 0 < z \lesssim 2 $. Therefore, using the present value $ H_0 $ \cite{pdg}
we obtain,
$$
 k_Q = 0.242  \; ({\rm Gpc})^{-1} \; .
$$
Notice that the value of $ k_Q $ is {\bf smaller} than the characteristic scale $ k_1 $.

\medskip

We display in Table IV the relevant wavenumbers: $ k_Q, \; k_1 , \; k_0$ 
 and the number of efolds since the beginning of inflation when they
exit the horizon. We see that the quadrupole modes {\bf exit} the horizon during the
fast-roll stage, approximately $ 1/10 $ of an efold before the end of fast roll.  The
mode $ k_1 $ exit the horizon by $ \ln a = 1.107 $, very close to the point $
\ln a = 1.091 $ where $ \epsilon_v = 1/N $.  That is, $ k_1 $ precisely exits
the horizon when {\bf fast roll ends and becomes slow roll}.

\begin{table}
  \begin{tabular}{c|c|c|}
    $~~~~~ k ~~~~~ $    & $ \ln a $ at horizon exit 
    &  $ \epsilon_v $  at horizon exit \\ \hline
    $ k_Q = 0.242 \; {\rm Gpc}^{-1} $ &  $ 1.01  $ & $ 0.0276 \gtrsim 1/N $  \\ \hline
    $  k_1 = 0.266  \; {\rm Gpc}^{-1} $ & $ 1.107 $ & $ 0.0188 \sim 1/N $  \\ \hline
    $  k_{0} = 2 \; {\rm Gpc}^{-1} $ (WMAP)  &  $ 3.135 $ 
    & $ \lesssim 1/N $  \\ \hline
    $  k_{0} = 50 \; {\rm Gpc}^{-1} $ (CosmoMC)  &  $ 6.363 $ 
    & $ \lesssim 1/N $  \\ \hline
  \end{tabular}   \;.
\caption{The number of efolds since the beginning of inflation when the
  wavenumbers $ k_Q, \; k_1 , \; k_{0} $ exit the horizon. 
The quadrupole modes {\bf exit} the horizon during the fast roll stage, 
about 
$ 1/10 $ of an efold before fast roll ends. $ k_1 $ 
precisely exits the horizon at the {\bf transition} 
from the fast roll to the slow roll stage.}
\end{table}

\medskip

We denote by $ k_{0} $ in Table IV the pivot wavenumbers 
in the WMAP \cite{WMAP-3} and CosmoMC codes \cite{lewis},
 where the indices 
$ n_s , \; r $ and the running of $ n_s $ are computed. Both $ k_{0}$'s  
exit the horizon well inside the slow roll regime.

\medskip

We read from Table IV that the {\bf total} number of efolds of inflation is given
by 
$$
 N_{tot} = N + 6  
$$
since we have six efolds before the  pivot wavenumber in CosmoMC
exit the horizon followed by $ N $ efolds of inflation.

\medskip

We can compute the redshift $ 1 + z_b $ since the begining of inflation till today
comparing  $ k_Q  = 0.242  \; ({\rm Gpc})^{-1} $ (today) with
$ k_Q^{initial}  = 0.910 \; k_1^{initial} = 12.7 \; m $ 
(at the begining of inflation).
[Recall that $ 1 \, {\rm GeV} = 1.564 \times  10^{41} \; ({\rm Gpc})^{-1} $].

We use for $ m $ the value obtained from the scale of inflation \cite{mcmc}
$$
m = \frac{M^2}{M_{Pl}}
$$
where $ M $ is fixed by the amplitude of the scalar adiabatic fluctuations \cite{WMAP-3,mcmc}.
We obtain 
$$
 M = 0.57 \times 10^{16} \;  {\rm GeV} \quad {\rm and}  \quad
 m = 1.34 \times 10^{13} \; {\rm GeV ~~~ for}  \quad   y = 1.322  \; . 
$$
[Notice that these results are in agreement with the generic estimates eq.(\ref{m}).]
Therefore, 
$$
 1 + z_b = 1.10 \times 10^{56} \simeq e^{129} \; .
$$
Assuming a sharp transition from inflation to radiation dominated expansion,
the redshift $ 1 + z_b $ can be written as
\be\label{29}
 10^{-56} \sim \frac1{1 + z_b} = a_r \; e^{-N_{tot}} \; ,
\ee
where $  a_r $ is the scale factor at the begining of the radiation dominated era
and $ N_{tot} $ is the total number of efolds during inflation (during fast-roll plus
during slow-roll).

The scale factor at the beginning ($ a_r $) and the end ($ a_{eq} $, equilibration) of the 
radiation dominated era are related by
$$
\frac{a_{eq}}{a_r} = \sqrt{\frac{H}{H_{eq}}} \; .
$$
where $ H $ and $ H_{eq} $ stand for the Hubble parameter at the beginning and at the end
of the radiation dominated era, respectively. For simplicity we assume instantaneous
reheating in these formulas.

Furthermore, $ H_{eq} $ and $ H_0 $ are related by \cite{mass}
$$
H_{eq} = \sqrt{2 \; \Omega_m} \; a_{eq}^{-\frac32} \; H_0 \; .
$$
where $ H_0 $ stands for the Hubble parameter today and $ \Omega_m $ for the matter fraction
of the energy density of the universe today.

Using the current values of the cosmological parameters \cite{pdg} we find
\be\label{ar}
a_r \sim  10^{-29} \; \sqrt{\frac{10^{-4} \; M_{Pl}}{H}} 
\simeq e^{-67} \; \sqrt{\frac{10^{-4} \; M_{Pl}}{H}} \; .
\ee
Inserting eq.(\ref{ar}) into eq.(\ref{29}) yields
\be\label{H57}
\frac{H}{10^{14} \; {\rm GeV}} = e^{2[56 - N]} \; .
\ee
Since $ H $ must be below its value at the beginning of inflation $ \sim 10^{14} $ GeV
[see eq.(\ref{Hi})], we conclude that
\be\label{cots}
  N > 56 \; .
\ee
On the other hand, we know from BBN (Big Bang Nucleosynthesis) that $ H $ is at least 
larger than 1 MeV. This together with eq.(\ref{H57}) yields the upper bound
\be\label{coti}
  N < 76 \; .
\ee
Furthermore, our MCMC simulations give good fits for $ N \sim 50-60 $. 
The bound eq.(\ref{cots})
therefore favours $ N \sim 60 $ which implies $ N_{tot} \sim 66 $ and 
$ H \sim 3 \times 10^{10} $ GeV. In addition, from eqs.(\ref{cots}) 
and (\ref{coti}) we obtain the bounds $ 62 < N_{tot} < 82 $.

\medskip

In summary, the fast-roll stage explains the quadrupole suppression and {\bf fixes the total
number of efolds} of inflation \cite{quadru1,quadru2}.

Our present MCMC analysis yields  $ N_{tot} \sim 66 $. More generally, the  upper bound
eq.(\ref{coti}) implies $ N_{tot} < 82 $.

Changing $ N $ from 50 to 60 does not affect significatively the MCMC fits we present in this paper. 
This is partially due to the fact that a change on $ y $ can partially compensate a change on $ N $.
Another {\bf hint} to increase  $ N $ above 50 comes from WMAP-5 that gives a larger $ n_s $ and using
the theoretical upper limit for $ n_s $ \cite{prd,mcmc}:
\be\label{cotns}
n_s < 1 - \frac{1.9236\ldots}{N} \; ,
\ee
which gives $ n_s < 0.9679 \ldots $ for $ N = 60 $.
This value is compatible with the $ n_s $ value from 
WMAP$5+$BAO$+$SN and no running \cite{WMAP-5}.

\begin{acknowledgments}
We thank L. Page, D. Boyanovsky, A. Hajian, R. Rebolo and D. Spergel for fruitful discussions.
The MCMC simulations were performed on the Turing and Atena Linux Clusters  of the
Physics Department G. Occhialini of the University Milano-Bicocca.  
\end{acknowledgments}

\section{Appendix}

As already shown in sec. \ref{mcmcsec}, the four types of correlation
multipoles, TT, TE, EE and BB, among the CMB anisotropies can be numerically computed with good
accuracy within the CosmoMC program starting from a given cosmological model
with a fixed value for all its parameters. To be precise, this computation is
performed by the CAMB subprogram, which is an evolution of CMBFAST 
\cite{selzal}. CAMB can also compute, with several levels
of approximations, the matter power spectrum observable today given the
primordial power spectrum of density perturbations.

\medskip

On modern workstation CPU's, the calculation of two thousand scalar multipoles
(related to the primordial curvature fluctuations) and one thousand tensor
multipoles (related to primordial gravitational waves) takes less than one
second. Thus, within a given type of cosmological model, the theoretical
predictions for different choices of parameters can be produced at a very high
rate.

To test these predictions against the experimental data CosmoMC makes use of
likelihood functions to assign different weights to different sets of
correlation multipoles and matter power data. The experimental data and
associated numerical code to evaluate such likelihoods is part of CosmoMC in the
case of small--scale CMB experiments and LSS surveys. The data and likelihood
routines for WMAP are not part of CosmoMC, but they can be downloaded from 
http://lambda.gsfc.nasa.gov/ and integrated quite easily into CosmoMC, since 
its interface to CMB likelihoods has been appositely designed for WMAP.

The WMAP likelihood code is particularly complex with respect to the other experiments,
due to the wealth of experimental data, the variety of source of systematical
errors and the importance of cosmic variance on lower multipoles. Indeed, it has
significally changed and improved along the three WMAP releases. At any rate,
it is used in CosmoMC exactly as released by the WMAP team.

\medskip

Finally, CosmoMC provides the MCMC engine, that is, the routines to perform
suitable random walks (the chains) in the parameter space of a give cosmological
model in such a way to reconstructs the experimental probability for the
parameters from the distribution of values produced along the chains. In our
simulations we always employed the default Metropolis rule, where the one-step
transition probability from one set $\lambda$ of parameter values to the next is
given by
\begin{equation*}
  W(\lambda',\lambda) = g(\lambda',\lambda)
  \; {\rm min} \, \left\{1\,,\; \frac{L(\lambda') \;
      g(\lambda',\lambda)}{L(\lambda) \; g(\lambda,\lambda')} \right\}
\end{equation*}
where $ g(\lambda',\lambda) $ is a Gaussian proposal, or jump, probability and
$ L(\lambda) $ is the complete posterior likelihood, that is the product of the
prior probability of choosing the starting point of the chain, times the
likelihood obtained by comparing the theoretical prediction on multipoles (and
matter power if LSS constraints are required) with the experimental data. As is well
known from the theory of stationary Markov chains, in the limit of infinitely
long chains no dependence is retained on $g(\lambda',\lambda)$ and the
reconstructed profile is that of $L(\lambda)$ only. Of course, since actual
chains have a finite length, suitable convergence tests are needed to verify
that such a reconstruction is accurate enough.  On pc-cluster running several
parallel chains at once, CosmoMC offers very effective tests of this kind.

\medskip

In order to test our own models based on Binomial New Inflation, with or without
sharp--cut or fast-roll, we needed to modify some routines in CosmoMC. Since our
changes with respect to the $\Lambda$CDM model are restricted to the primordial power
spectrum, only routines relative to the so--called fast variables needed suitable
modifications. Fast variables in CosmoMC are cosmological parameters that affect
only the primordial spectrum so that, in a Monte Carlo step that proposes
changes restricted to them, no need arises to perform the time--consuming
recomputation of the transfer functions from a given primordial spectrum to
observable multipoles. Slow variables such as the baryonic and dark matter
fractions, $\omega_b, \; \omega_c$, the optical depth $ \tau $, and the present
Hubble parameter $ H_0 $ have the opposite definition. No change was done
in the by far major portion of the CosmoMC program that deals with slow
variables. 

We recall also that we let only the four slow variables mentioned
above vary in our MCMC run (to be precise, we used the default choice of CosmoMC
which replaces $H_0$ with $ \theta $, the ratio of the approximate sound horizon
to the angular diameter distance), keeping all other slow cosmological
parameters, such as the parameter of the dark energy equation of state or the
neutrino density fraction, to the values of the standard $\Lambda$CDM model.

\end{document}